\documentclass[aps,prd,tightenlines,nofootinbib,reprint,superscriptaddress]{revtex4-1}

\usepackage{graphicx}
\usepackage{dcolumn}
\usepackage{bm}
\usepackage{amsmath}
\usepackage{multirow}

\begin{document}

\newcommand{\mevcc}{\!\mathrm{MeV}\!/c^2}
\newcommand{\mevc}{\!\mathrm{MeV}/\!c}
\newcommand{\mev}{\!\mathrm{MeV}}
\newcommand{\gevcc}{\!\mathrm{GeV}/\!c^2}
\newcommand{\gevc}{\!\mathrm{GeV}/\!c}
\newcommand{\gev}{\!\mathrm{GeV}}

\title{Hyperon Form Factors \& Diquark Correlations}

\author{S.~Dobbs}
\author{Kamal~K.~Seth}
\author{A.~Tomaradze}
\author{T.~Xiao}
\affiliation{Northwestern University, Evanston, Illinois 60208, USA}
\author{G. Bonvicini}
\affiliation{Wayne State University, Detroit, Michigan 48202, USA}

\date{\today}

\begin{abstract}
Using $e^+e^-$ annihilation data taken at the CESR collider with the CLEO-c detector, measurements of hyperon pair production cross sections and  elastic and transition electromagnetic form factors have been made at the charmonium resonances:  
$\psi(2S)$, $\sqrt{s}=3.69$~GeV, $|Q^2|=13.6$~GeV$^2$, $\mathcal{L}=48$~pb$^{-1}$; 
$\psi(3770)$, $\sqrt{s}=3.77$~GeV, $|Q^2|=14.2$~GeV$^2$, $\mathcal{L}=805$~pb$^{-1}$; 
and $\psi(4170)$, $\sqrt{s}=4.17$~GeV, $|Q^2|=17.4$~GeV$^2$, $\mathcal{L}=586$~pb$^{-1}$.  
Results with good statistical precision are obtained with high efficiency particle identification.
Systematics of pair production cross sections, and form factors with respect to the number of strange quarks in the hyperons are studied, and
evidence is presented for effects of diquark correlations in comparative results for $\Lambda^0$ and $\Sigma^0$, both of which have the same $uds$ quark content.
\end{abstract}

\maketitle

\section{Introduction}

The Universe is made of baryons.  With six different species of quarks, ground state baryons can be made in 20 different combinations of three quarks.  Of these twenty, only one, the proton is stable, and is available as a target for the study of its structure by means of scattering experiments in which spacelike ($Q^2$ positive) momentum transfer (four-momentum$^2\equiv\,$three-momentum$^2-t^2$) is made, leading to electromagnetic spacelike form factors.  This has led to extensive studies of the structure of the proton~\cite{slff}.  In contrast, studies of the structure of other baryons can only be made by production experiments for timelike ($Q^2$~negative) momentum transfers.  

Although the importance of studying hyperon structure, and measurement of timelike form factors of hyperons, was pointed out as early as 1960 by Cabibbo and Gatto~\cite{cabibbo}, experimental measurements became possible only with the advent of $p\bar{p}$ and $e^+e^-$ colliders, and the first measurements were reported only thirty years later.  In 1990, DM2 Collaboration at Orsay reported the first measurement of the production of $\Lambda^0$ and $\Sigma^0$ and their timelike form factors~\cite{dm2}, and in 2007 the BaBar Collaboration at SLAC reported~\cite{babar} measurement of elastic form factors of $\Lambda^0$, $\Sigma^0$ and $\Lambda^0\Sigma^0$ transition form factors using the ISR technique.  Both the DM2 and BaBar measurements were made near threshold energies, and very few counts were observed.  Small statistics and small momentum transfer (generally $<5$~GeV$^2$) did not lend these measurements to interpretation in terms of pQCD.  

The first measurements of hyperon pair production at large momentum transfer were made by the CLEO Collaboration at Cornell in 2005.  They reported branching fractions for the production of $\Lambda$, $\Sigma$, and $\Xi$ hyperons at the $\psi(2S,~3686~\text{MeV})$ resonance for $|Q|^2=13.6$~GeV$^2$~\cite{cleo2s}.  It was subsequently noted that pQCD predicts that, unlike at $\psi(2S)$, resonance production of hadron pairs at $\psi(3770)$ and $\psi(4170)$ was expected to be very small, and non-resonance electromagnetic production of hadron pairs would dominate, and it could be used to determine electromagnetic form factors for large timelike momentum transfers. 
We use the pQCD prediction that the hadronic and leptonic decays of $\psi(nS)$ states scale similarly with the principal quantum number $n$, i.e.,
\begin{multline}
\frac{\mathcal{B}(\psi(n'S)\to\text{gluons}\to\text{hadrons})}{\mathcal{B}(\psi(nS)\to\text{gluons}\to\text{hadrons})} \\  = \frac{\mathcal{B}(\psi(n'S)\to\gamma^*\to\text{electrons})}{\mathcal{B}(\psi(nS)\to\gamma^*\to\text{electrons})},
\end{multline}
to estimate that the resonance contribution to data taken at the $\psi(3770)$ and $\psi(4170)$ is negligibly small, and these data can be used to determine timelike form factors of hadrons.
The validity of this expectation was confirmed by us in successful measurements  of the form factors of pion, kaon, and proton at $\psi(3770)$ and $\psi(4170)$~\cite{cleo-ff}. 
Using the measured branching fractions for the $J/\psi$ and $\psi(2S)$~\cite{pdg}, and the present luminosities and efficiencies, we determine that the expected number of events is $3.0~\Lambda^0$, $1.4~\Sigma^+$, $1.2~\Sigma^0$, $1.2~\Xi^-$, $0.6~\Xi^0$, and $0.3~\Omega^-$ for resonance decays of the $\psi(3770)$ in the present measurements, and $2.0~\Lambda^0$, $1.0~\Sigma^+$, $0.9~\Sigma^0$, $0.9~\Xi^-$, $0.4~\Xi^0$, and $0.2~\Omega^-$ for resonance decays of the $\psi(4170)$.  In other words, the contributions of resonance decays are negligibly small in all cases, and the observed events at $\psi(3770)$ and $\psi(4170)$ can be safely attributed to electromagnetic production, $e^+e^- \to \gamma^* \to B\overline{B}$, and can be used to determine form factors.

Using this assumption we made measurements of timelike form factors of $\Lambda^0$, $\Sigma^0$, $\Sigma^+$, $\Xi^0$, $\Xi^-$, $\Omega^-$ hyperons for $|Q^2|=14.2$~GeV$^2$ and 17.4~GeV$^2$, and reported our first results in 2014~\cite{hyperonff}.  Since then, we have substantially improved (by factors of $3-5$) the efficiency of our hyperon identification, and in this paper we present our final results for the electromagnetic form factors of hyperons with improved precision.
We also present for the first time our results for the $\Lambda^0\Sigma^0$ transition form factor, and we update our results for pair production cross sections and branching fractions for $\psi(2S)$ decay.

\begin{figure*}[!tb]
\begin{center}
\includegraphics[width=5.5in]{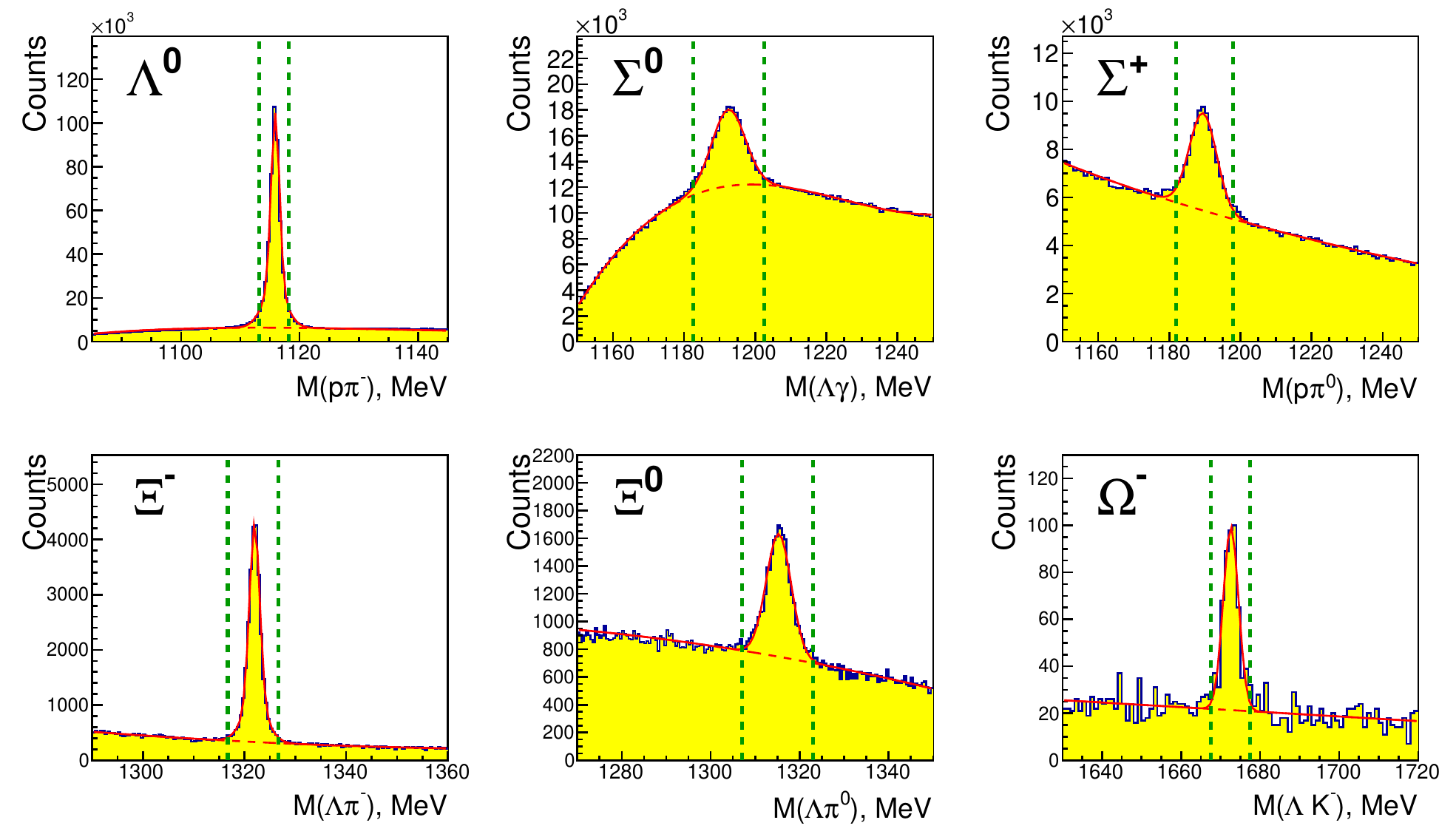}

\end{center}
\caption{Invariant mass distributions for $\psi(2S)$ data.  The solid red curves show the results of the fits to these spectra, while the dashed red line shows the background component of the fit.  Clear peaks corresponding to each hyperon are seen, and their fitted yields are displayed in each panel.  The dashed vertical line correspond to the ``signal'' region used for the momentum plots in Fig.~2.}
\label{fig:psi2sinclmass}
\end{figure*}

\begin{figure*}[!tb]
\begin{center}
\includegraphics[width=5.5in]{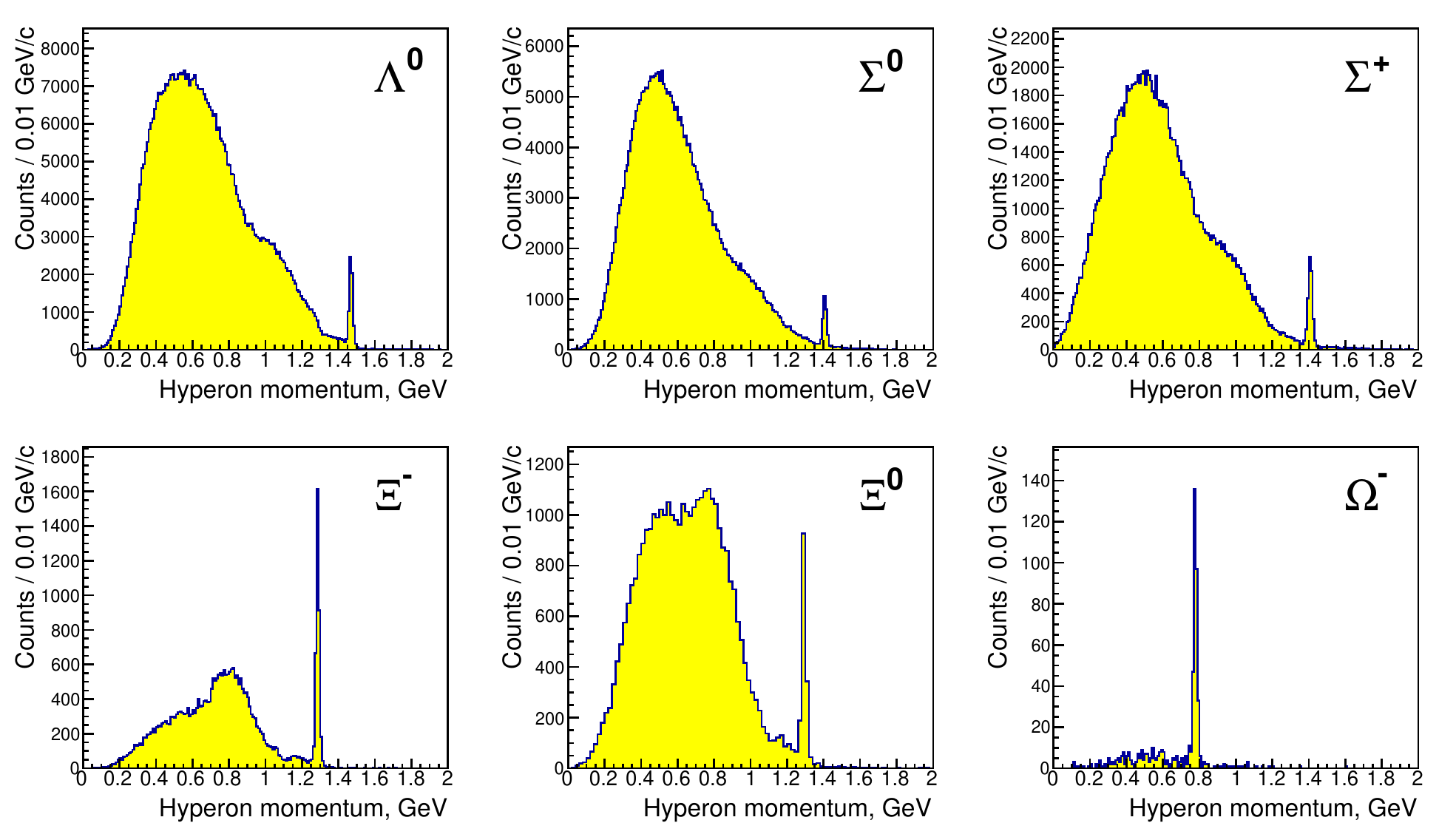}

\end{center}
\caption{Momentum distributions for hyperon candidates in the ``signal'' mass regions defined in Fig.~1 for $\psi(2S)$ data.  The clear peaks at high momentum are due to pair-production of hyperons.  The yields at lower momenta are due to hyperons produced in association with other hadrons and the combinatorial backgrounds underneath the hyperon peaks seen in Fig.~1.}
\label{fig:psi2smom}
\end{figure*}

\begin{figure*}[!tb]
\begin{center}
\includegraphics[width=5.5in]{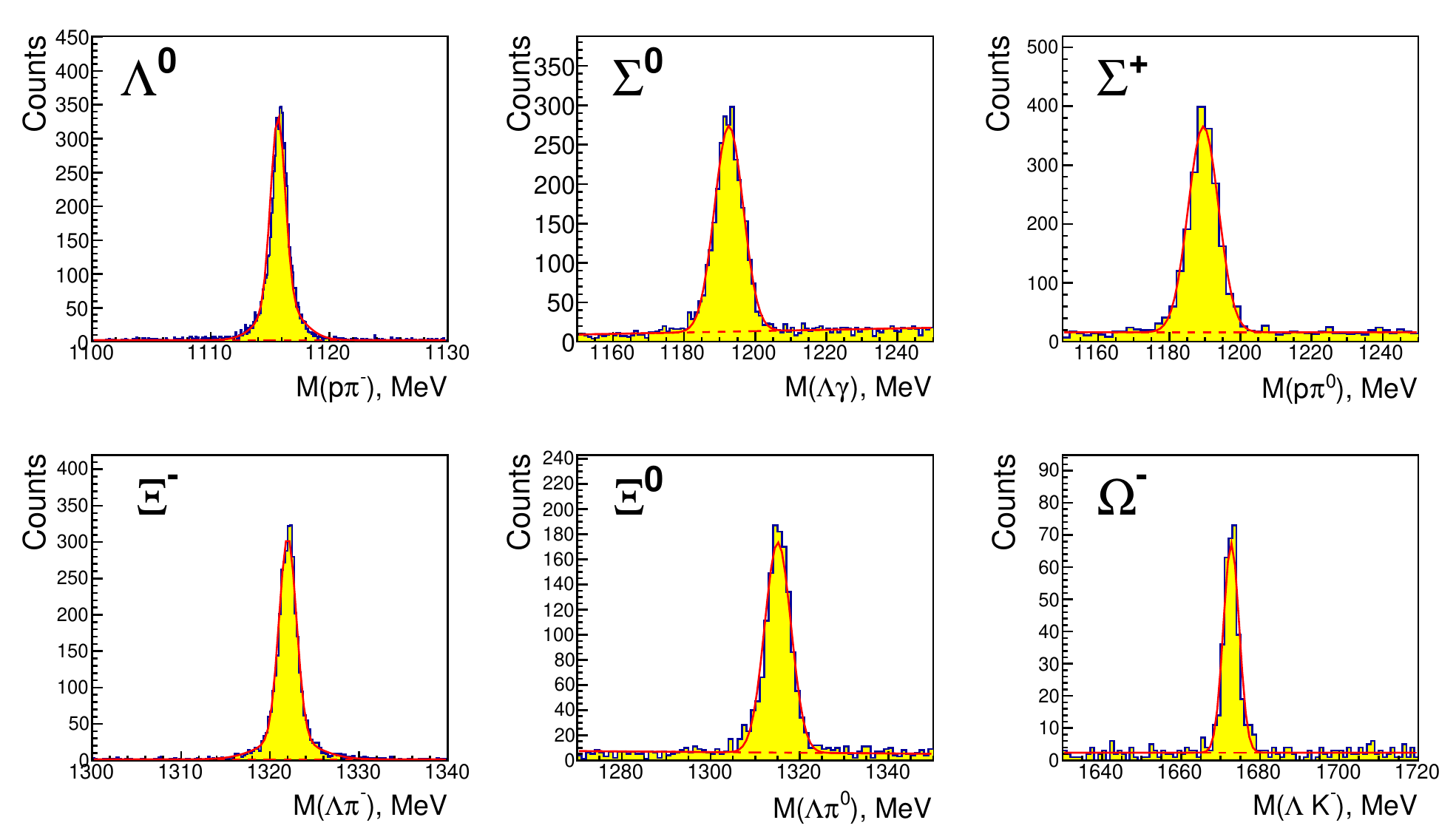}

\end{center}
\caption{Invariant mass distributions for hyperon candidates in $\psi(2S)$ data in the pair-production region given by $E(B)/E(\text{beam})=0.99-1.01$. The solid red curves show the result of the fit to this spectrum described in the text, while the dashed red line shows the background component of the fit.}
\label{fig:psi2smass}
\end{figure*}

\section{Data Samples and Event Selections}

We use data taken with the CLEO-c detector, which has been described in detail elsewhere~\cite{cleodetector}.
The data were taken at $\psi(3686)$, $\sqrt{s}=3.69$~GeV, $\psi(3770)$, $\sqrt{s}=3.77$~GeV, $\psi(4170)$, $\sqrt{s}=4.17$~GeV, with integrated luminosities of $\mathcal{L}=48$~pb$^{-1}$, 805~pb$^{-1}$, and 586~pb$^{-1}$ at $\sqrt{s}=3.69$~GeV, 3.77~GeV, and 4.17~GeV, respectively.  
We identify hyperons by their principal decay modes~\cite{pdg}: $\Lambda^0\to p\pi^-$ (63.9\%), $\Sigma^+\to p\pi^0$ (51.6\%), $\Sigma^0 \to \Lambda^0 \gamma$ (100\%), $\Xi^-\to\Lambda^0\pi^-$ (99.9\%), $\Xi^0\to\Lambda^0\pi^0$ (99.5\%), $\Omega^- \to \Lambda^0 K^-$ (67.8\%) [charge conjugate decay modes are included].  We note that in all but $\Sigma^+$, a $\Lambda^0$ is produced which leads to a displaced vertex and very clean hyperon identification.  The event selections used to reconstruct these hyperon decays are similar to those described in our previous publication~\cite{hyperonff}, and are briefly described below.

Charged particles ($\pi^\pm$, $K^\pm$, $p/\bar{p}$) are required to have $|\cos\theta|<0.93$, where $\theta$ is the polar angle with respect to the $e^+$ beam.  To identify charged particles, we use the combined likelihood variable
\begin{small}
$$\Delta \mathcal{L}_{i,j} = [-2\ln L^\mathrm{RICH} + (\chi^{dE/dx})^2]_i - [-2\ln L^\mathrm{RICH} + (\chi^{dE/dx})^2]_j,$$
\end{small}
where $i,j$ are the particle hypotheses $\pi,K,p$, $dE/dx$ is the measured energy loss in the drift chamber, and $L^\mathrm{RICH}$ is the log-likelihood of the particle hypothesis using information from the RICH detector.
We identify protons by requiring that the measured properties of the charged particle be more like a proton than either a charged pion or kaon by $3\sigma$, i.e., $\Delta \mathcal{L}_{p,\pi}<-9$ and $\Delta \mathcal{L}_{p,K}<-9$.  Kaons from the decay $\Omega^-\to \Lambda^0 K^-$ suffer from larger backgrounds, and a stricter requirement of $\Delta \mathcal{L}_{K,\pi}<-25$ and $\Delta \mathcal{L}_{K,p}<-25$ is used.

Any number of photons are allowed in an event.  Photon candidates are calorimeter showers in the ``good barrel'' ($|\cos\theta|=0-0.81$) or ``good endcap'' ($|\cos\theta|=0.85-0.93$) regions that do not contain one of the few noisy calorimeter cells, are inconsistent with the projection of a charged particle track, and have a transverse energy deposition consistent with that of an electromagnetic shower.  We reconstruct $\pi^0\to\gamma\gamma$ decays by requiring that photon candidate pairs have mass within $3\sigma$ of the known $M(\pi^0)$, and then kinematically fitting them to $M(\pi^0)$.  The $\pi^0$ candidates are initially assumed to originate from the interaction point, however the $\pi^0$ candidates used to reconstruct $\Sigma^+$ and $\Xi^0$ candidates are refit with the assumption that they originate at the decay vertex of their primary hyperon.

We identify hyperons by kinematically fitting them under the assumption that all particles originate from a common vertex, and require that this vertex be displaced from the interaction point by $>3\sigma$.  
The $\Lambda^0$ hyperons are reconstructed by combining two oppositely charged tracks.  The higher momentum track is required to be identified as a negative proton, and the lower momentum track is assumed to be a negative pion.  
When reconstructing hyperons which decay into a $\Lambda^0$, each $\Lambda^0$ candidate is further required to be consistent with its nominal mass of $M(\Lambda^0)=1115.683$~MeV~\cite{pdg} within $5\sigma$. It is then kinematically fitted to this nominal mass, and is required to have a decay vertex at a greater distance from the interaction point than that of the hyperons decaying into $\Lambda^0$.

The $\Sigma^+$ hyperons are reconstructed by combining protons with $\pi^0$ candidates.  
Only $\Sigma^+$ candidates with a kinematic fit $\chi^2$ of $<20$ are kept.

The $\Sigma^0$ hyperons are reconstructed by combining a $\Lambda^0$ candidate with a photon candidate. The photon candidate is required to have an energy greater than 50~MeV.  

The $\Xi^-$ and $\Omega^-$ hyperons are reconstructed by combining a $\Lambda^0$ candidate with a charged track identified as $\pi^-$ and $K^-$, respectively. 

The $\Xi^0$ hyperons are reconstructed similarly to the $\Sigma^+$ hyperon, with the proton replaced by a $\Lambda^0$ candidate, and an additional requirement of the kinematic fit $\chi^2<20$.


\begin{figure*}[!tb]
\begin{center}
\includegraphics[width=5.5in]{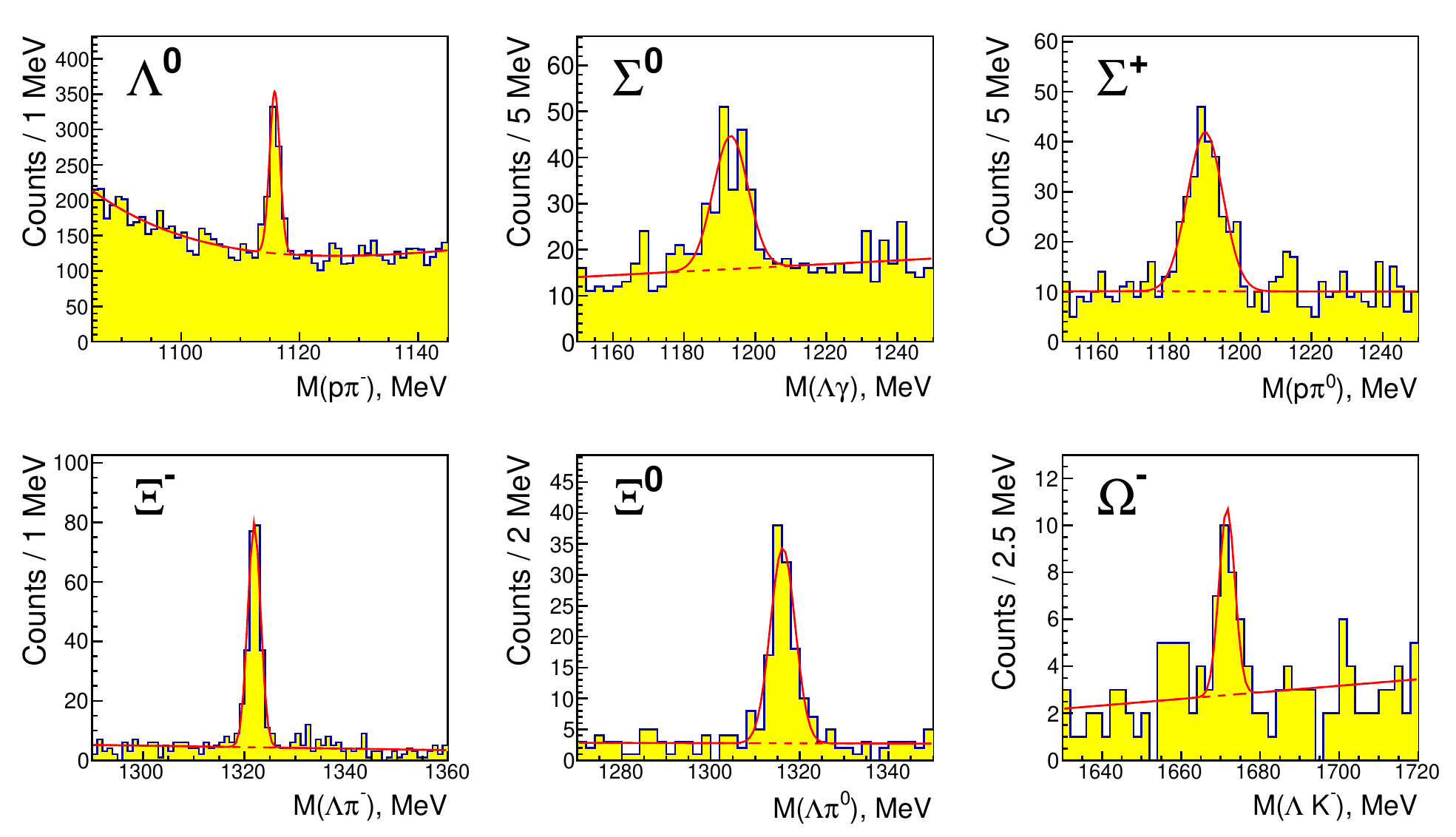}

\end{center}
\caption{Invariant mass distributions for hyperon candidates in $\psi(3770)$ data in the pair-production region given by $E(B)/E(\text{beam})=0.99-1.01$. The solid red curves show the result of the fit to this spectrum described in the text, while the dashed red line shows the background component of the fit.}
\label{fig:3770mass}
\end{figure*}

\begin{figure*}[!tb]
\begin{center}
\includegraphics[width=5.5in]{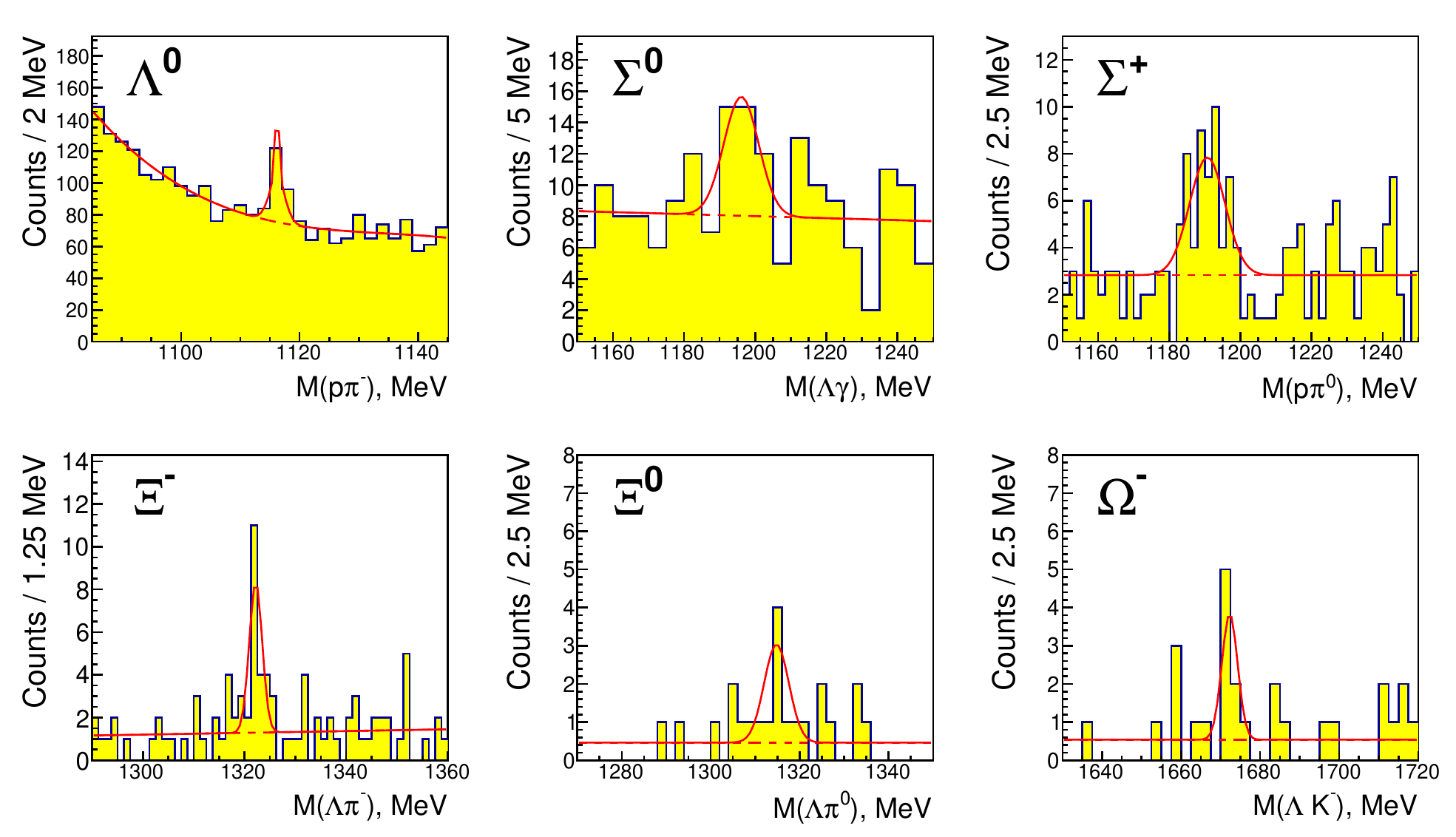}

\end{center}
\caption{Invariant mass distributions for hyperon candidates in $\psi(4170)$ data in the pair-production region given by $E(B)/E(\text{beam})=0.99-1.01$. The solid red curves show the result of the fit to this spectrum described in the text, while the dashed red line shows the background component of the fit.}
\label{fig:4170mass}
\end{figure*}

\begin{table*}[!tb]
\caption{Summary of cross section and branching fraction results from $\psi(2S)$ data.  The systematic uncertainties are taken from Table~\ref{tbl:systematics}.  Note that the results for protons are borrowed from our Ref.~\cite{cleo-ff}.  The uncertainties in our present results are smaller than our results in  Ref.~\cite{cleo-ff} by factors two or larger.}
\begin{ruledtabular}
\begin{tabular}{l|c@{\hspace{10pt}}|c@{\hspace{10pt}}|ccc@{\hspace{10pt}}|cc}
$B$ & $N_\text{fit}^{\psi(2S)}$ & $N_\text{ff}$ & $\epsilon_B$ (\%) & $\sigma_B$ (pb) & $\mathcal{B}\times10^4$ & $\mathcal{B}\times10^4$ (prev.) \cite{cleo-ff} & BES-III~\cite{bes1} \\
\hline

$p$ & $4475\pm78$ & $16\pm10$ & $63.1$ & $196\pm3\pm12$ & $3.08\pm0.05\pm0.18$ & --- & --- \\

\hline

$\Lambda^0$ & $6531\pm82$ & $42\pm3$ & $71.6$ & $244.7\pm3.1\pm10.1$ & $3.71\pm0.05\pm0.15$ & $3.75\pm0.09\pm0.23$ & $3.97\pm0.02\pm0.12$ \\ 
$\Sigma^0$  & $2645\pm56$ & $14\pm2$ & $48.6$ & $145.6\pm3.1\pm7.1$ & $2.22\pm0.05\pm0.11$ & $2.25\pm0.11\pm0.16$ & $2.44\pm0.03\pm0.11$ \\  
$\Sigma^+$  & $1874\pm46$ & $15\pm1$ & $33.0$ & $151.4\pm3.8\pm6.4$ & $2.31\pm0.06\pm0.10$ & $2.51\pm0.15\pm0.16$ & --- \\
$\Xi^-$     & $3580\pm61$ & $17\pm1$ & $48.2$ & $199.9\pm3.4\pm9.4$ & $3.03\pm0.05\pm0.14$ & $2.66\pm0.12\pm0.20$ & $2.78\pm0.05\pm0.14$ \\  
$\Xi^0$     & $1242\pm38$ & $ 8\pm1$ & $25.6$ & $131.6\pm4.1\pm7.1$ & $1.97\pm0.06\pm0.11$ & $2.02\pm0.19\pm0.15$ & --- \\
$\Omega^-$  &  $326\pm19$ & $ 1\pm1$ & $25.8$ & $ 33.7\pm2.0\pm2.0$ & $0.52\pm0.03\pm0.03$ & $0.47\pm0.09\pm0.05$ & --- \\
\hline
$\Lambda^0\Sigma^0$  & $30\pm5$ & $0.2\pm0.1$ & $9.9$ & $8.1\pm1.5\pm0.5$ & $0.123\pm0.023\pm0.008$ & ---  & ---  \\
\end{tabular}

\end{ruledtabular}
\label{tbl:psi2sresults}
\end{table*}

\begin{table*}[!tb]
\caption{Summary of cross section and form factor results from $\psi(3770)$ data. The systematic uncertainties are taken from Table~\ref{tbl:systematics}.   The results for protons are borrowed from our Ref.~\cite{cleo-ff}.  The cross sections $\sigma_B(\text{BES-III})$ are calculated from the results in Ref.~\cite{bes3} assuming $\mathcal{L}(\text{BES-III})=2.9$~fb$^{-1}$ and $C(\text{BES-III})=0.8$.  Note that the electromagnetic $\sigma_B$ in column 3 are generally smaller than the resonance decay cross sections from $\psi(2S)$ in Table~I by orders of magnitude.}
\begin{ruledtabular}

\begin{tabular}{l|ccc@{\hspace{10pt}}|c@{\hspace{10pt}}|c@{\hspace{10pt}}|c}
$B$ & $N_\text{fit}^{\psi(3770)}$ & $\epsilon_B$ (\%) & $\sigma_B$ (pb)  & $\sigma_B(\text{BES-III})$ (pb) \cite{bes3} & $G_M\times10^2$ & $G_M\times10^2$ (prev.)~\cite{cleo-ff}\\
\hline

$p$ & $215\pm15$ & $71.3$ & $0.46\pm0.03\pm0.03$ & --- & $0.88\pm0.03\pm0.02$ & --- \\

\hline
$\Lambda^0$ & $498\pm39$ & $74.8$ & $1.08\pm0.09\pm0.04$ & ---  & $1.48\pm0.06\pm0.03$ & $1.18\pm0.06\pm0.04$\\
$\Sigma^0$  & $142\pm20$ & $48.0$ & $0.48\pm0.07\pm0.02$ & $0.26\pm0.04\pm0.02$ & $1.01\pm0.07\pm0.02$ & $0.71\pm0.09\pm0.03$ \\
$\Sigma^+$  & $200\pm19$ & $32.3$ & $1.02\pm0.10\pm0.04$ & $0.82\pm0.10\pm0.07$ & $1.47\pm0.07\pm0.03$ & $1.32\pm0.13\pm0.04$ \\
$\Xi^-$     & $240\pm17$ & $55.0$ & $0.71\pm0.05\pm0.03$ & $0.48\pm0.07\pm0.04$ & $1.28\pm0.04\pm0.03$ & $1.14\pm0.09\pm0.04$ \\
$\Xi^0$     & $111\pm12$ & $24.6$ & $0.71\pm0.08\pm0.03$ & $0.80\pm0.12\pm0.06$ & $1.28\pm0.07\pm0.03$ & $0.81\pm0.21\pm0.03$ \\
$\Omega^-$  & $ 20\pm 6$ & $29.5$ & $0.11\pm0.03\pm0.01$ & ---  & $0.63\pm0.09\pm0.02$ & $0.64^{+0.21}_{-0.25}\pm0.03$\\

\hline
$\Lambda^0\Sigma^0$  & $29\pm5$ & 10.8 & $0.43\pm0.08\pm0.03$ & --- & $0.77\pm0.07\pm0.03$ & --- \\

\end{tabular}

\end{ruledtabular}
\label{tbl:3770results}
\end{table*}

\begin{table*}[!tb]
\caption{Summary of cross section and form factor results from $\psi(4170)$ data.  The systematic uncertainties are taken from Table~\ref{tbl:systematics}.  The results for protons are borrowed from our Ref.~\cite{cleo-ff}.  Note that the $\sigma_B$ in column 3 for hyperon pair production at $\psi(4170)$ are smaller by factors 4 to 10 than these for $\psi(3770)$ in Table~II.}
\begin{ruledtabular}
\begin{tabular}{l|ccccc}
$B$ & $N_\text{fit}^{\psi(4170)}$ & $\epsilon_B$ (\%) & $\sigma_B$ (pb) & $G_M\times10^2$ & $|Q^4|G_M[3770]/|Q^4|G_M[4170]$ \\
\hline

$p$ & $92\pm10$ & $68.7$ & $0.29\pm0.03\pm0.02$ & $0.76\pm0.04\pm0.02$ & $0.77\pm0.05$ \\

\hline
$\Lambda^0$ & $65\pm15$ & $64.9$ & $0.23\pm0.05\pm0.01$ & $0.73\pm0.08\pm0.02$ & $1.28\pm0.16$ \\
$\Sigma^0$  & $19\pm 7$ & $46.0$ & $0.09\pm0.04\pm0.02$ & $0.47\pm0.09\pm0.04$ & $1.23\pm0.27$ \\
$\Sigma^+$  & $31\pm 8$ & $30.7$ & $0.23\pm0.06\pm0.04$ & $0.75\pm0.09\pm0.06$ & $1.16\pm0.18$ \\ 
$\Xi^-$     & $18\pm 5$ & $53.2$ & $0.08\pm0.02\pm0.01$ & $0.44\pm0.06\pm0.01$ & $1.80\pm0.25$ \\
$\Xi^0$     & $ 7\pm 3$ & $25.8$ & $0.06\pm0.03\pm0.01$ & $0.40\pm0.08\pm0.04$ & $1.89\pm0.41$ \\
$\Omega^-$  & $ 7\pm 3$ & $33.7$ & $0.04\pm0.02\pm0.01$ & $0.39\pm0.08\pm0.01$ & $0.92\pm0.23$ \\

\hline 
$\Lambda^0\Sigma^0$  & $7.0^{+3.6}_{-2.9}$ & 10.8 & $0.15^{+0.07}_{-0.06}\pm0.01$ & $0.50^{+0.12}_{-0.09}\pm0.02$ & $1.02^{+0.27}_{-0.21}$ \\
\end{tabular}

\end{ruledtabular}
\label{tbl:4170results}
\end{table*}

\section{Results}

We present our results for pair production of hyperons from $\psi(2S)$ decays in Sec.~III.A, and our results for the determination of timelike form factors of hyperons for the data at $\psi(3770)$ and $\psi(4170)$ in Sec.~III.B.  We present our first results for the determination of the $\Lambda^0\Sigma^0$ transition form factor in Sec.~III.C.

\subsection{Resonance Production of $\bm{\Lambda}$, $\bm{\Sigma}$, $\bm{\Xi}$,  and $\bm{\Omega}$ Hyperons at $\bm{\psi(2S)}$}

In Fig.~\ref{fig:psi2sinclmass} we show the raw invariant mass spectra for the $\psi(2S)$ data as obtained by identifying either a single hyperon or antihyperon.

In Fig.~\ref{fig:psi2smom}, we show the momentum distributions for the hyperon candidates in the signal mass regions bounded by the dashed vertical lines in Fig.~\ref{fig:psi2sinclmass}.  The sharp peaks at high momenta in these distributions are due to pair production of hyperons $B\overline{B}$.  The large yields at lower momenta are due to hyperons produced in association with other hadrons (mostly pions and kaons), $B$ or $\overline{B}+X$, and combinatorial backgrounds underneath the hyperon peaks in Fig.~\ref{fig:psi2sinclmass}.

\begin{figure*}[!tb]
\begin{center}
\includegraphics[width=5.6in]{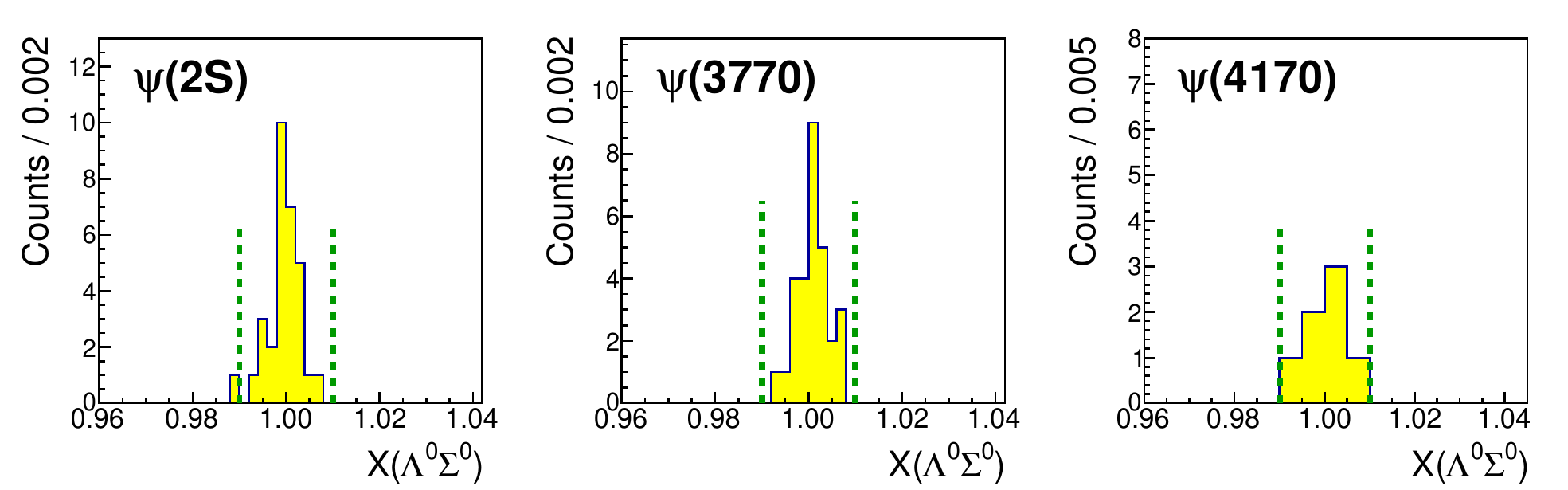}
\end{center}
\caption{ $\Lambda^0\Sigma^0$ yield distributions as function of $\text{X} \equiv \left[ E(\Lambda^0) + E(\Sigma^0) \right] / \sqrt{s}$.}
\label{fig:lamsig0}
\end{figure*}

The yield of the pair-produced hyperons can be conveniently obtained as the events which satisfy the requirement $[E(B)~\text{or}~E(\overline{B})]/E(\text{beam})=0.99-1.01$.  The invariant mass distributions of these events is shown in Fig.~\ref{fig:psi2smass}.  Simple fits to these spectra with small constant backgrounds lead to the results listed in Table~\ref{tbl:psi2sresults}.  From these fits, we obtain
\begin{equation}
\sigma_0[\psi(2S)] = \frac{ N_\text{signal} }{ \epsilon_B\,\mathcal{L} \, C},
\end{equation}
where $N_\text{signal} = N_\text{fit} - N_\text{ff}$, $\epsilon_B$ is the MC-determined efficiency, $\mathcal{L}=48~$pb$^{-1}$ is the $e^+e^-$ luminosity which leads to $N(\psi(2S)~\text{produced})=24.5\times10^6$, and $C$ is the radiative correction factor of $0.76-0.78$.  The contribution of form factor events in these data, $N_\text{ff}$, are estimated by pQCD-based extrapolations, assuming a $s^{-5}$ cross section dependence, from the timelike form factor measured in Sec.~III.B. at $\psi(3770)$.  The branching fractions are calculated as
\begin{equation}
\mathcal{B}(\psi(2S)\to B^+B^-) = \frac{ N_\text{signal} }{ \epsilon_B\, N[\psi(2S)]} .
\end{equation}
The numerical results are presented in Table~\ref{tbl:psi2sresults}.

For comparison, we also list in Table~\ref{tbl:psi2sresults} our earlier published results~\cite{cleo-ff}, as well as the recent results by BES-III for $\Lambda^0$, $\Sigma^0$, and $\Xi^-$ pair production.  We note that the BES-III results are in good agreement with ours.  Cross sections and branching fractions corresponding to the large yields for single hyperon\,+\,X inclusive production require evaluation of momentum-dependent efficiencies, and are not presented here, except to note that the ratio $\sigma(\Lambda^0)/\sigma(\Sigma^0)$ for the inclusive $\Lambda^0$ and $\Sigma^0$ production is found to be $4.1\pm0.6$.

\subsection{Form Factor Measurements}

The data for $\psi(3770)$ and $\psi(4170)$ are analyzed for hyperon pair production in exactly the same manner as the $\psi(2S)$ data.  The invariant mass spectra for $\psi(3770)$ are shown in Fig.~\ref{fig:3770mass}, and those for $\psi(4170)$ in Fig.~\ref{fig:4170mass}.  The numerical results are presented in Tables \ref{tbl:3770results} and \ref{tbl:4170results}.

As expected, the yields for electromagnetic production of hyperon pairs are much smaller than those for resonance production in the case of $\psi(2S)$, despite factors $10-20$ larger luminosities.  The MC efficiencies differ from those for $\psi(2S)$ only by small amounts.  The resulting pair production cross sections are smaller by factors as large as several hundred.

As can be seen in Table~\ref{tbl:4170results}, the yield of hyperon pair production at $\psi(4170)$ is smaller by factors 4 to 10 than that for $\psi(3770)$, and the cross sections have substantially larger errors, which lead to fits of poorer quality in Fig.~\ref{fig:4170mass}.  This is mainly due to differences in luminosity, and the fact that according to QCD quark counting rules~\cite{qc_rules}, baryon form factor cross sections fall as $s^{-5}$.

In Tables~\ref{tbl:3770results} and \ref{tbl:4170results}, we also show results for the determination of timelike form factors using the conventional relation between cross sections and electric and magnetic form factors $G_E(s)$ and $G_M(s)$ of spin$-1/2$ nucleons.

It has become conventional to analyze pair production cross sections for the determination of timelike form factors as is conventionally used to analyze cross sections for spacelike momentum transfers to determined spacelike form factors.  It is therefore instructive to review the relationship between the two.

Electromagnetic form factors are analytic functions of four-momentum transfer, $|Q^2|$.  It follows that form factors for timelike momentum transfer are related to those for spacelike momentum transfer by analytic continuation, and timelike and spacelike form factors they should be analyzed in the same formalism, i.e., in terms of the Dirac form factor $F_1$ and the Pauli form factor $F_2$, or equivalently, in terms of the electric form factor $G_E$ and the magnetic form factor $G_M$, with the relations $G_E=F_1+(s/m^2)F_2$ and $G_M=F_1+F_2$.  However, the physical meaning of $G_E$ and $G_M$ is not the same for spacelike and timelike momentum transfers.  While spacelike $G_E$ and $G_M$ are related to spatial distributions of charge and magnetic moment through Fourier transforms, timelike $G_E$ and $G_M$ are related to helicity correlations in the particle--antiparticle pair, with $F_2$ denoting photon coupling to particle--antiparticle pairs with parallel spins, and $F_1$ to pairs with antiparallel spins.

The relation between cross sections and $G_E$ and $G_M$ form factors for spin--1/2 hadrons is
\begin{equation}
\sigma_{B\overline{B}} = \left( \frac{4\pi\alpha^2\beta_B}{3s} \right) \left[ |G_M^B(s)|^2 + (2m_B^2/s) |G_E^B(s)|^2 \right]
\end{equation}
where $\alpha$ is the fine structure constant, $\beta_B$ is the velocity of the baryons in the center-of-mass system, and $m_B$ is the mass of the baryon $B$. 

Because the contributions of $G_E$ and $G_M$ terms have different angular dependences, it is possible to determine $|G_E/G_M|$ by analyzing the angular distributions of the cross sections. 
However, because of limited statistics it is generally not possible to determine $|G_E/G_M|$, and data are analyzed for two limiting values, $|G_E/G_M|=0$ and 1.

BaBar~\cite{babar} attempted to analyze their data for $\Lambda\bar{\Lambda}$ production in two different $\sqrt{s}$ bins assuming MC-determined modifications of the angular contributions of $G_E$ and $G_M$.
They obtained two quite different values, $|G_E/G_M|=1.73^{+0.99}_{-0.57}$ for the $\sqrt{s}=2.23-2.40$~GeV bin with 115 events, and $|G_E/G_M|=0.71^{+0.66}_{-0.71}$ for the $\sqrt{s}=2.40-2.80$~GeV bin with 61 events, but considered both of them as consistent with $|G_E/G_M|=1$, and analyzed their data with that assumption.



We have analyzed the angular distributions for our data for $\psi(3770)$, $Q^2=14.2~\mathrm{GeV^2}$, for three hyperons for which we have the largest number of events in Table II, $N(\Lambda^0)=498\pm39$, $N(\Xi^-)=240\pm17$, and $N(\Xi^0)=111\pm12$.
We follow the MC-based procedure described by BaBar, and for all three we obtain $|G_E/G_M|=0$, with $90\%$ confidence limits:
\begin{itemize}
\item $\Lambda^0$: $<0.17$
\item $\Xi^-$: $<0.32$
\item $\Xi^0$: $<0.29$
\end{itemize}
Our results for all three cases are thus consistent with $|G_E/G_M|=0$.

We therefore analyze our data assuming $G_E=0$.

We analyze $\Omega^-\overline{\Omega^-}$ cross section also using Eq.~(4), although, as noted by K\"orner and Kuroda, for spin--3/2 baryons the form factors includes higher-moment contributions~\cite{kornerkuroda}.

%

\subsection{$\Lambda^0\Sigma^0$ Transition Form Factor}

We use the reaction $e^+e^-\to\Lambda^0\Sigma^0$ to measure the $\Sigma^0\to\Lambda^0$ transition form factor, which requires us to reconstruct both the $\Lambda^0$ and $\Sigma^0$ separately. We also have to take into account that $\Sigma^0$ decays almost entirely via $\Sigma^0\to\gamma\Lambda^0$, with the transition photon of low energy ($\sim80$~MeV). 
To reconstruct the $\Lambda^0$ and $\Sigma^0$ for this reaction, we use the event selections as described before, except that protons are identified using the looser criteria of $\Delta \mathcal{L}_{p,\pi}<0$ and $\Delta \mathcal{L}_{p,K}<0$.  To select fully-reconstructed $\Lambda^0\overline{\Sigma^0}$ pairs, we require the total momentum of the $\Lambda^0\overline{\Sigma^0}$ pair to be less than 50~MeV.  To distinguish pair-produced $\Lambda^0\overline{\Sigma^0}$ candidates from $\Lambda^0\overline{\Sigma^0}$ candidates which come from $\Sigma^0\overline{\Sigma^0}$ events in which one of the  $\Sigma^0\to\gamma\Lambda^0$ transition photons is lost or ignored, we require the total momentum of the $\Lambda^0\overline{\Sigma^0}$ to be smaller than that of any $\Sigma^0\overline{\Sigma^0}$ pair in the event.

\begin{table}[!tb]
\caption{Summary of systematic uncertainties.  The total systematic uncertainty listed in the sum in quadrature of the individual contributions.}
\begin{ruledtabular}
\begin{tabular}{l|cccccc}
$\psi(2S)$ branching fractions              & $\Lambda^0$ & $\Sigma^0$ & $\Sigma^+$ & $\Xi^-$ & $\Xi^0$  & $\Omega^-$ \\
\hline 
$N(\psi(2S))$                 & 2 & 2 & 2 & 2 & 2 & 2 \\
Track reconstruction         & 2 & 2 & 1 & 3 & 2 & 3 \\
Particle ID                  & 2 & 2 & 2 & 2 & 2 & 4 \\
$\pi^0/\gamma$ reconstruction & 0 & 2 & 2 & 0 & 2 & 0 \\
Hyperon reconstruction       & 2 & 2 & 2 & 2 & 2 & 2 \\
Peak fitting                 & 1 & 2 & 1 & 1 & 3 & 1 \\
\textbf{$\bm{\psi(2S)}$ Total}               & \textbf{4.1} & \textbf{4.9} & \textbf{4.2} & \textbf{4.7} & \textbf{5.4} & \textbf{5.8} \\
\hline

Data              & $\Lambda^0$ & $\Sigma^0$ & $\Sigma^+$ & $\Xi^-$ & $\Xi^0$  & $\Omega^-$ \\
\hline 
Luminosity                   & 1 & 1 & 1 & 1 & 1 & 1 \\
Track reconstruction         & 2 & 2 & 1 & 3 & 2 & 3 \\
Particle ID                  & 2 & 2 & 2 & 2 & 2 & 4 \\
$\pi^0/\gamma$ reconstruction & 0 & 2 & 2 & 0 & 2 & 0 \\
Hyperon reconstruction       & 2 & 2 & 2 & 2 & 2 & 2 \\
Radiative corrections        & 0.2 & 0.2 & 0.2 & 0.2 & 0.2 & 0.2 \\
\hline
$\psi(3770)/\psi(4170)$ Common & 3.6 & 4.1 & 3.7 & 4.2 & 4.1 & 5.5 \\
\hline
$\psi(3770)$ Peak fitting                 & 2 & 5 & 3 & 3 & 1 & 8 \\
\textbf{$\bm{\psi(3770)}$ Total}               & \textbf{4.1} & \textbf{6.5} & \textbf{4.8} & \textbf{5.2} & \textbf{4.2} & \textbf{9.7} \\

\hline
$\psi(4170)$ Peak fitting                 & 5 & 16 & 17 & 2 & 18 & 5 \\
\textbf{$\bm{\psi(4170)}$ Total}               & \textbf{6.2} & \textbf{16.5} & \textbf{17.4} & \textbf{4.7} & \textbf{18.5} & \textbf{7.4} \\

\end{tabular}
 \end{ruledtabular}
\label{tbl:systematics}
\end{table}

Finally, the $\Lambda^0\overline{\Sigma^0}$ pair is kinematically fitted to the initial energy and momentum of the $e^+e^-$ collision, and the fit is required to have $\chi^2<20$.  If there are multiple $\Lambda^0\overline{\Sigma^0}$ candidate pairs in the event, the pair with the smallest $\chi^2$ is kept.  With these selection criteria, Monte Carlo studies show negligible backgrounds from the $\Lambda^0\overline{\Lambda^0}$ and $\Sigma^0\overline{\Sigma^0}$ final states.

The distribution of $\text{X}(\Lambda^0\Sigma^0) \equiv [E(\Lambda^0) + E(\Sigma^0)]/\sqrt{s}$ for each data set is shown in Fig.~\ref{fig:lamsig0}.  Clear peaks are seen in each case with essentially no background.  We take signal events to be in the range $\text{X}(\Lambda^0\Sigma^0) = 0.99 - 1.01$.  The $\Lambda^0\Sigma^0$ results for branching fractions, cross sections, and the form factors are calculated as previously described, and are summarized in the bottom rows of Tables~\ref{tbl:psi2sresults}, \ref{tbl:3770results}, and \ref{tbl:4170results}.   The systematic uncertainty in these branching fraction and cross section measurements is determined as described in Ref.~\cite{hyperonff}, and is found to be 6.7\%. 

\begin{figure}[!tb]
\begin{center}
\includegraphics[width=2.7in]{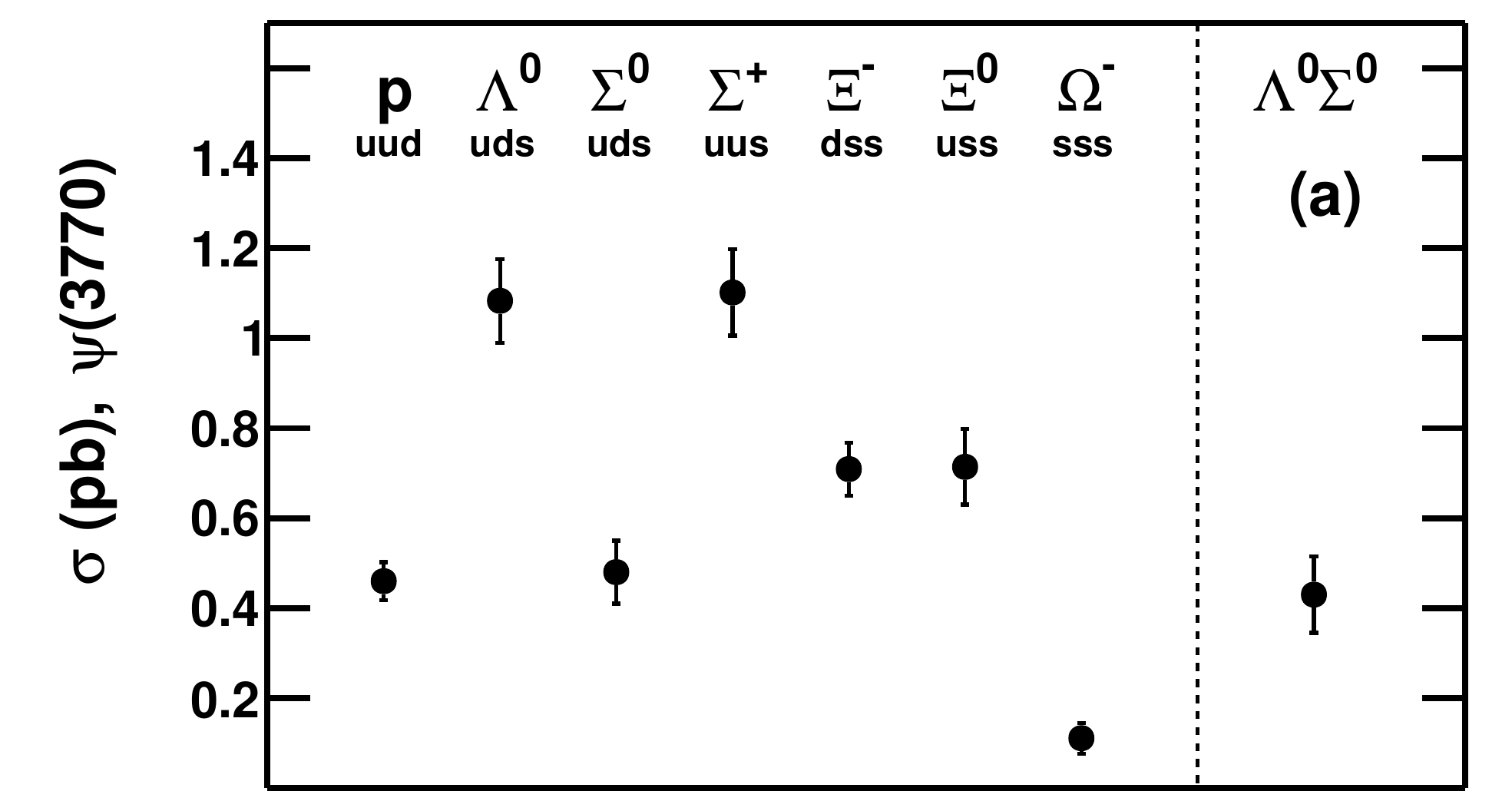}
\includegraphics[width=2.7in]{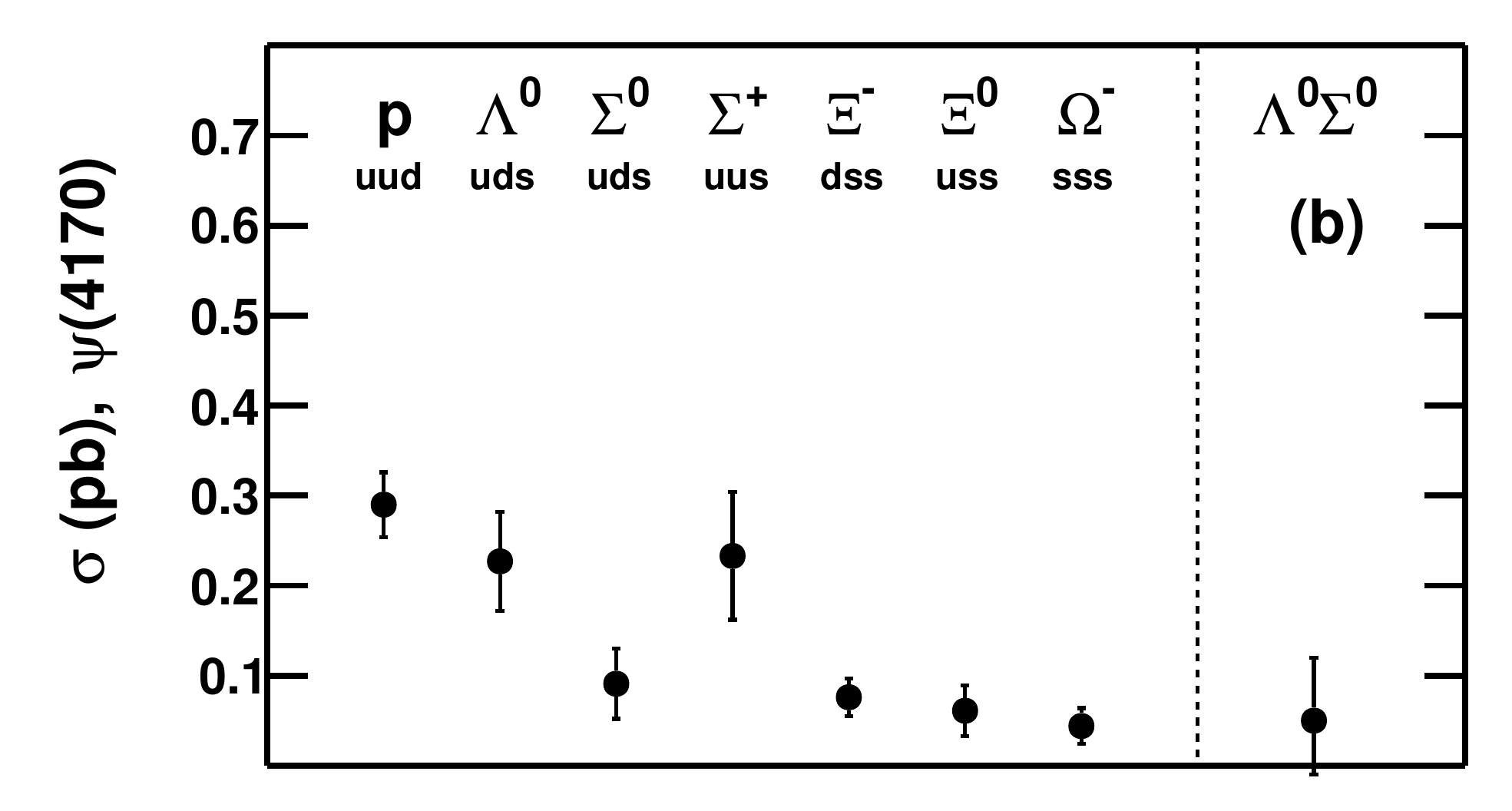}
\includegraphics[width=2.7in]{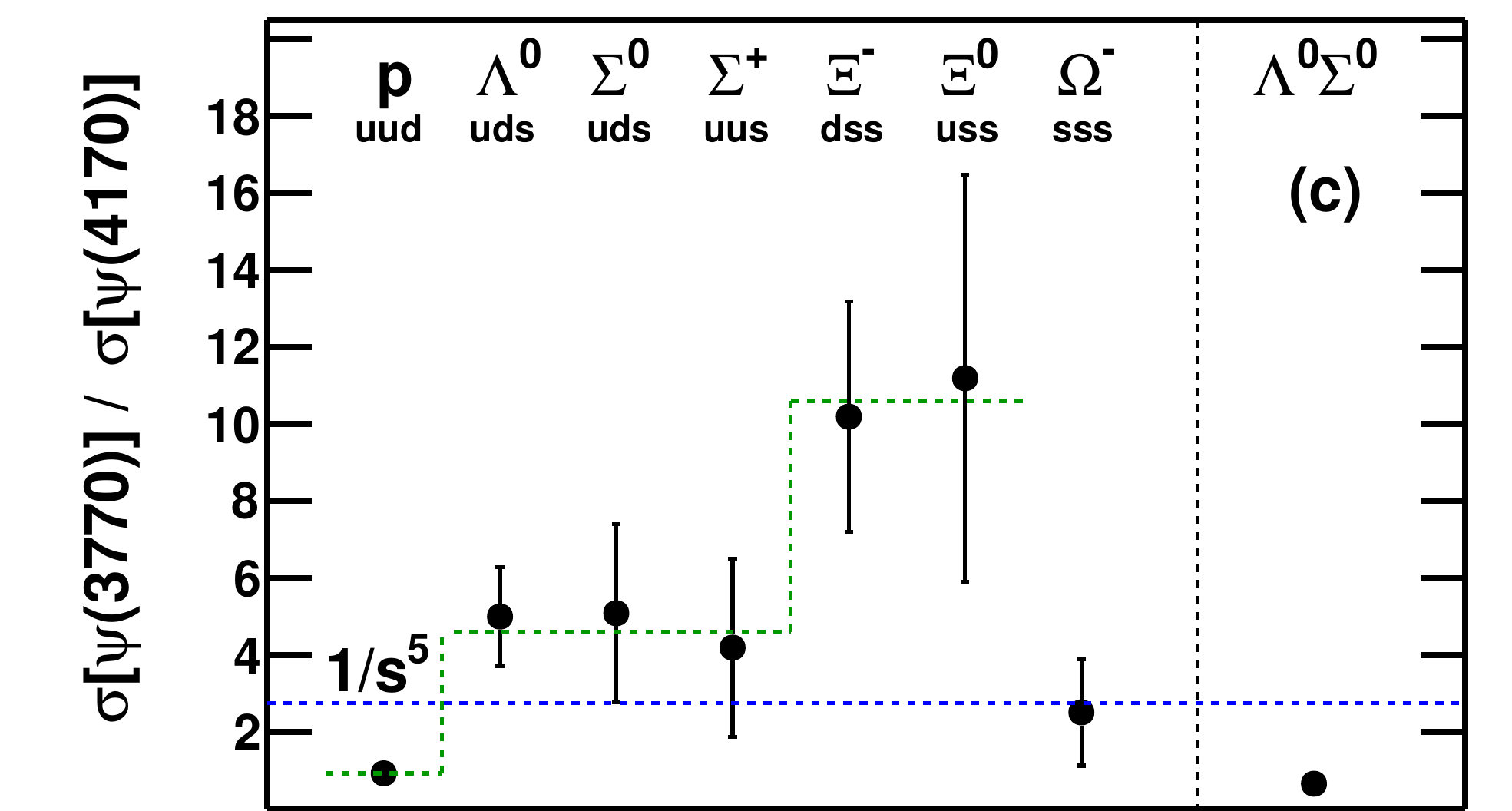}

\end{center}
\caption{Summary of cross section results.  The theoretical prediction for the ratios in panel (c) is $1/s^5=2.74$.  The data show systematics with clear differences between baryons containing 0, 1, or 2 strange quarks.}
\end{figure}

\begin{figure}[!tb]
\begin{center}
\includegraphics[width=2.7in]{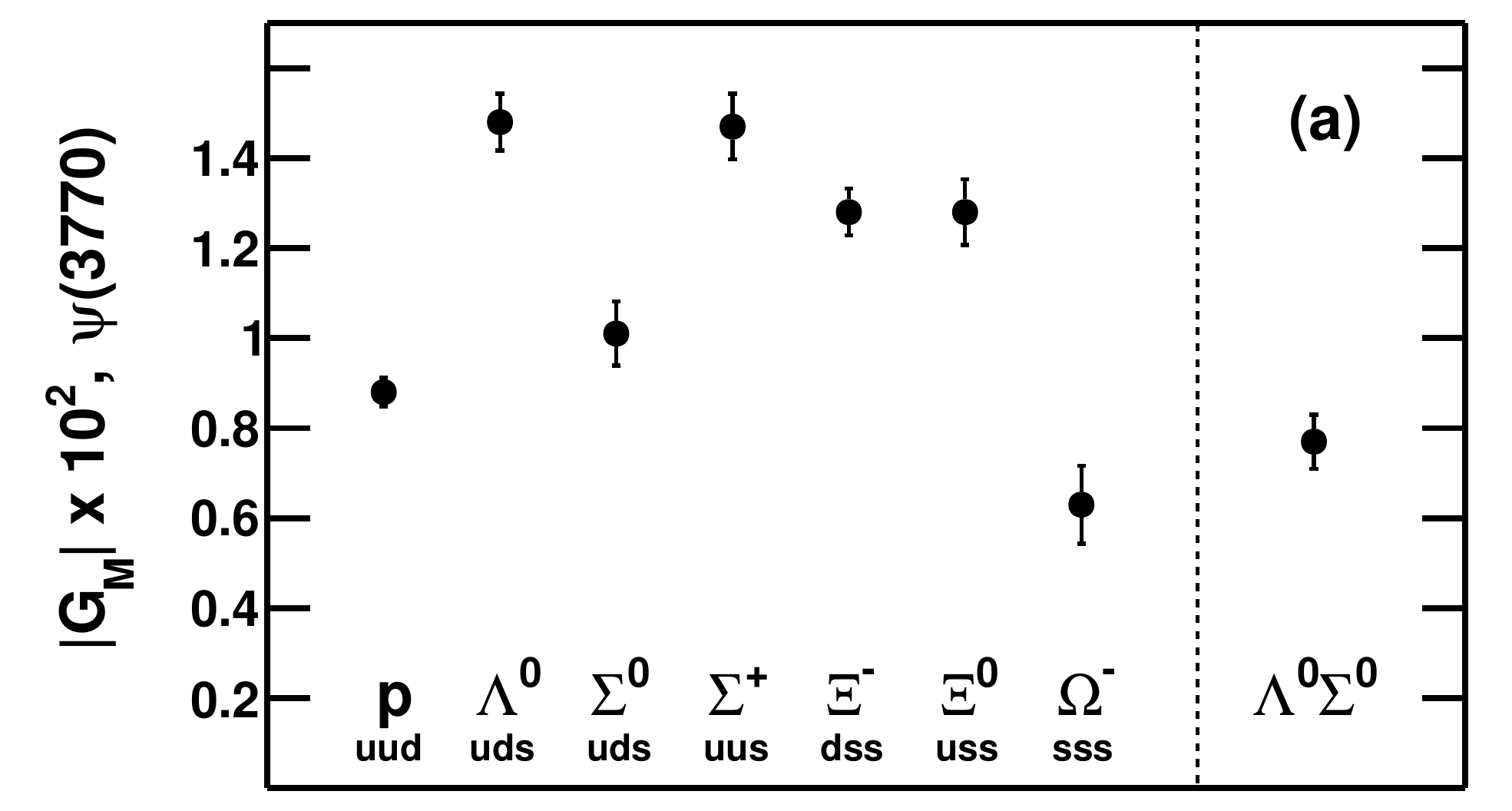}
\includegraphics[width=2.7in]{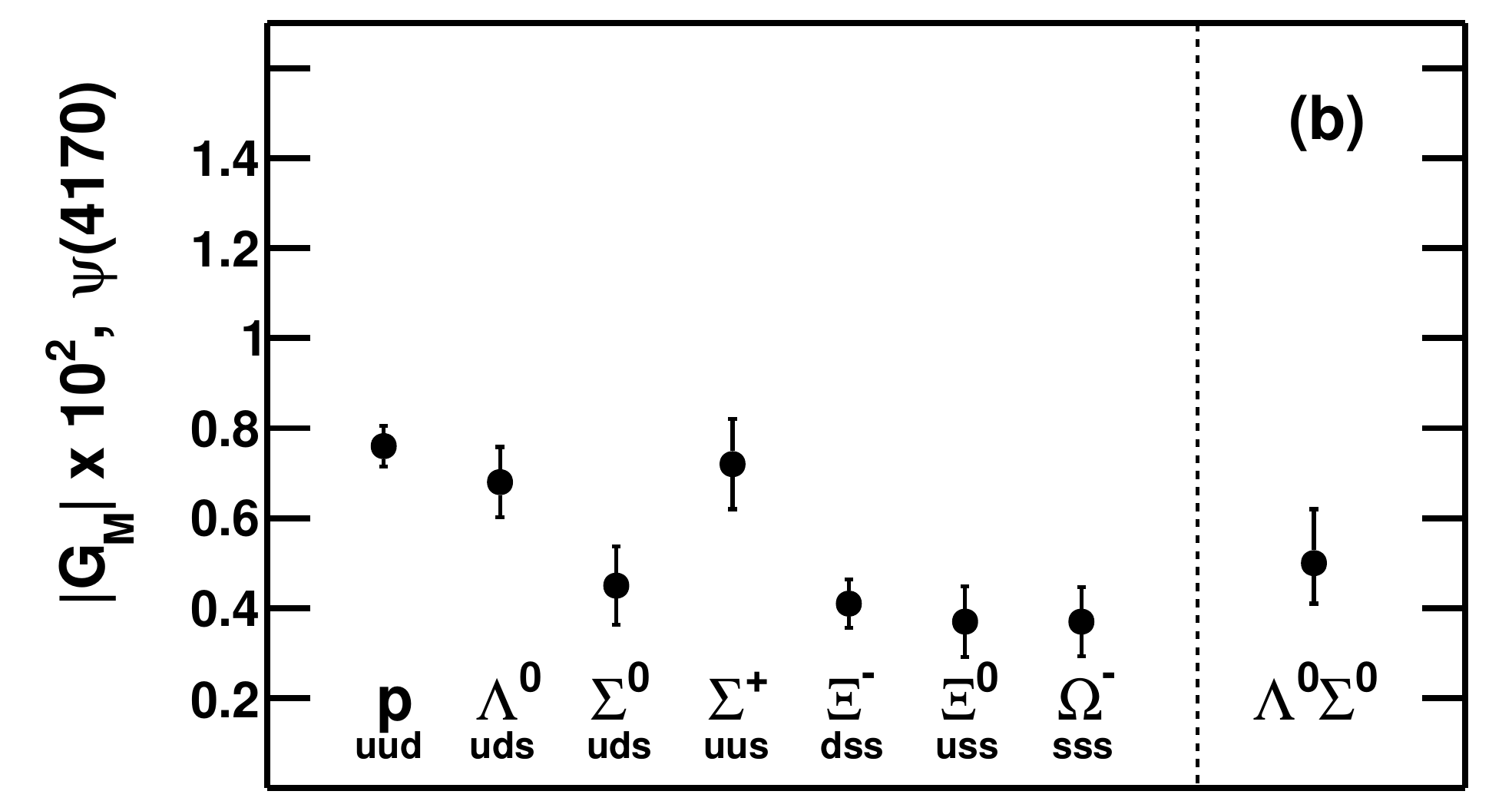}
\includegraphics[width=2.7in]{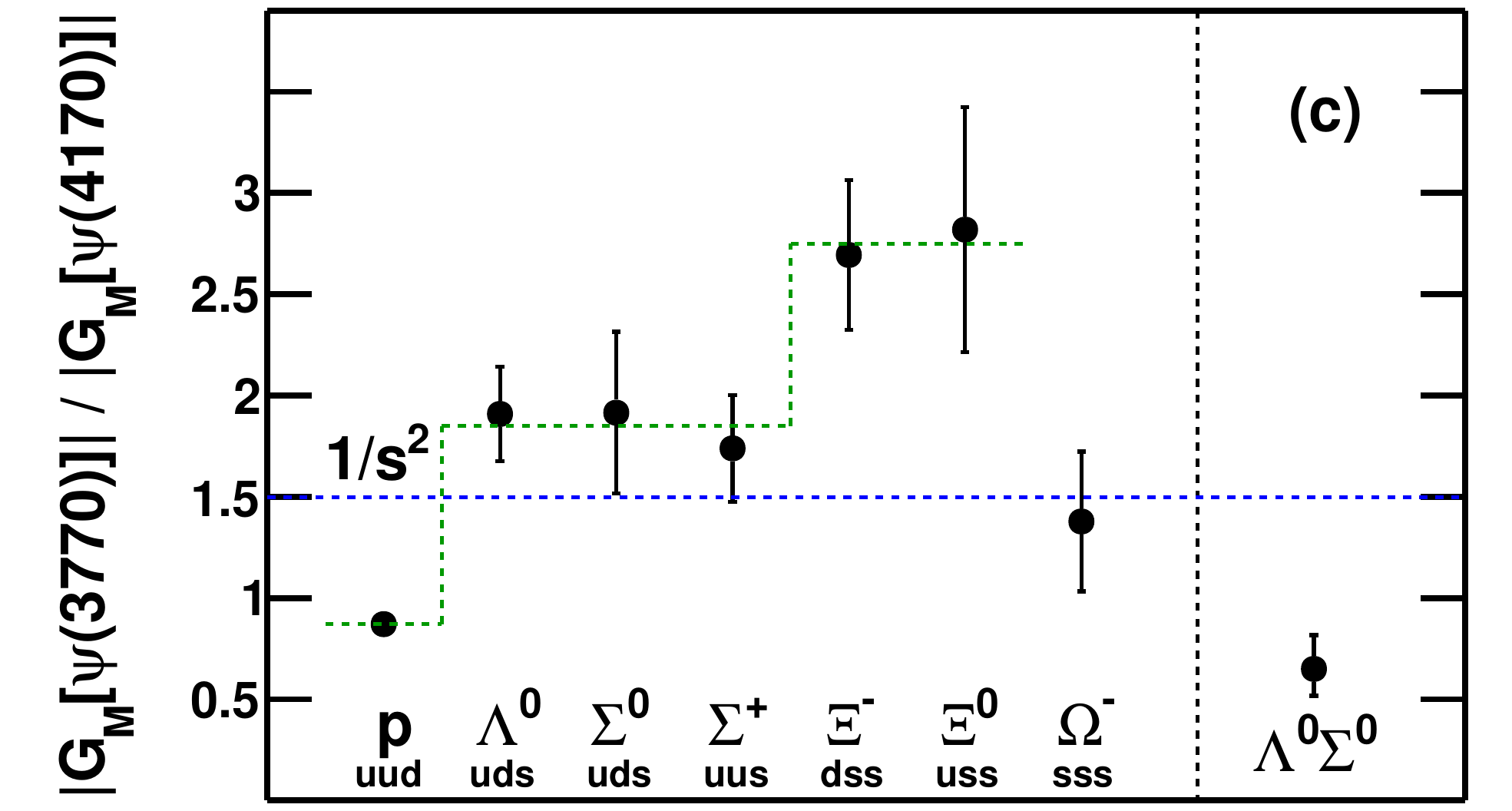}

\end{center}
\caption{Summary of form factor results.  The ratios of $G_M$ at $\psi(3770)$ and $\psi(4170)$ in panel (c) are predicted to be equal to $1/s^2=1.5$.  As noted for the cross section ratios in Fig.~7, the data above show different values for baryons containing 0, 1, and 2 strange quarks.}  
\end{figure}

\section{Systematic Uncertainties}

We evaluate systematic uncertainties due to various sources for each final state and add the contributions from the different sources together in quadrature.  The uncertainties due to particle reconstruction are 1\% per charged particle, 2\% per $\gamma$, 2\% per $\pi^0$, and 1\% per hyperon.  There are additional uncertainties of 2\% per $p$ and $K$ due to the use of RICH and $dE/dx$ information.  Other systematic uncertainties are 2\% in $N(\psi(2S))$, 1\% in $e^+e^-$ luminosity, and 0.2\% in the radiative corrections.  Uncertainties in hyperon peak fitting are evaluated by varying the order of the polynomial background and the fit range.  The largest variation of these is taken as the estimate of systematic uncertainty in peak fitting.  The individual values and quadrature sums are given in Table~\ref{tbl:systematics}.

\begin{figure*}[!tb]
\begin{center}
\includegraphics[width=1.9in]{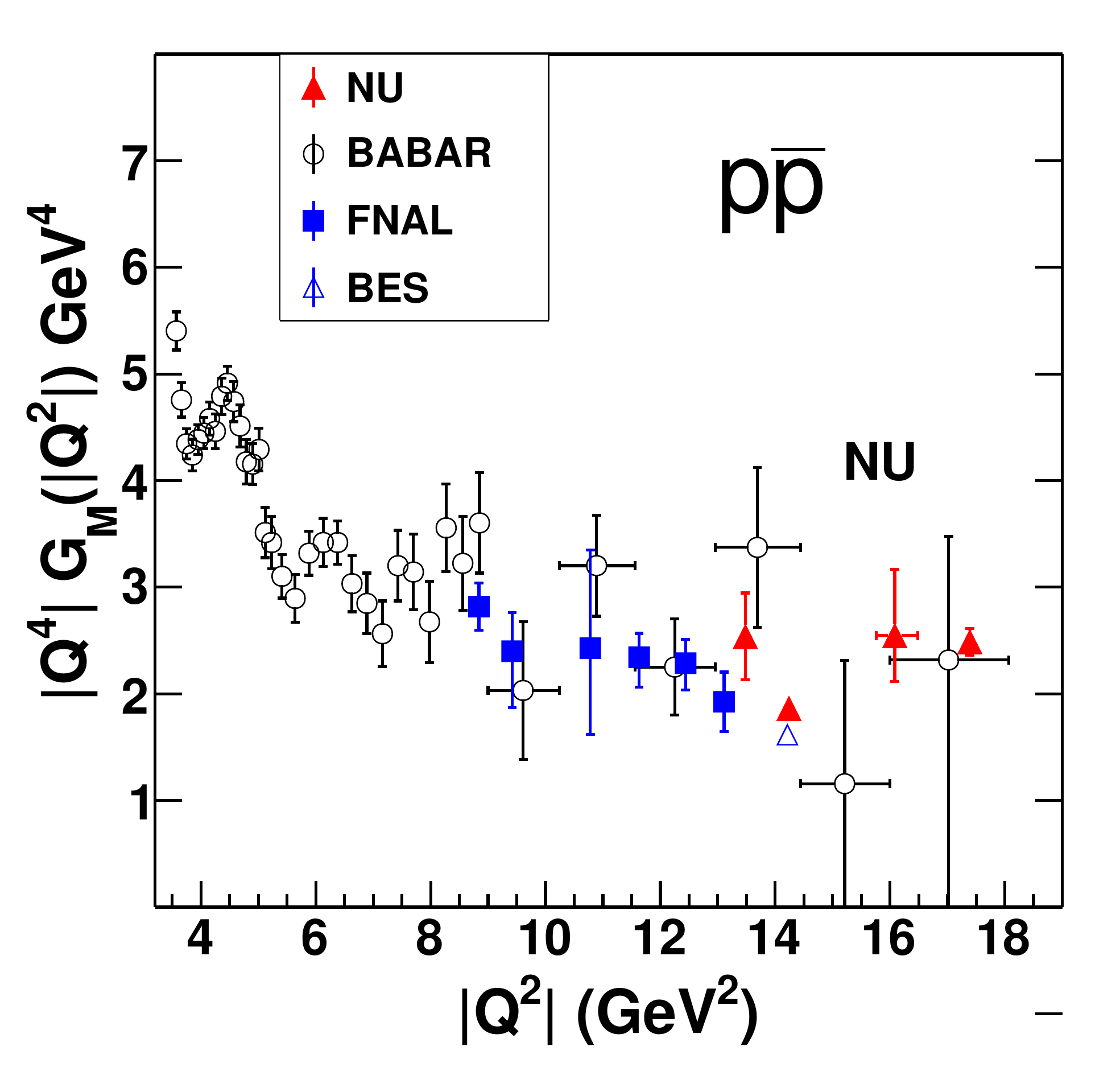}
\includegraphics[width=1.9in]{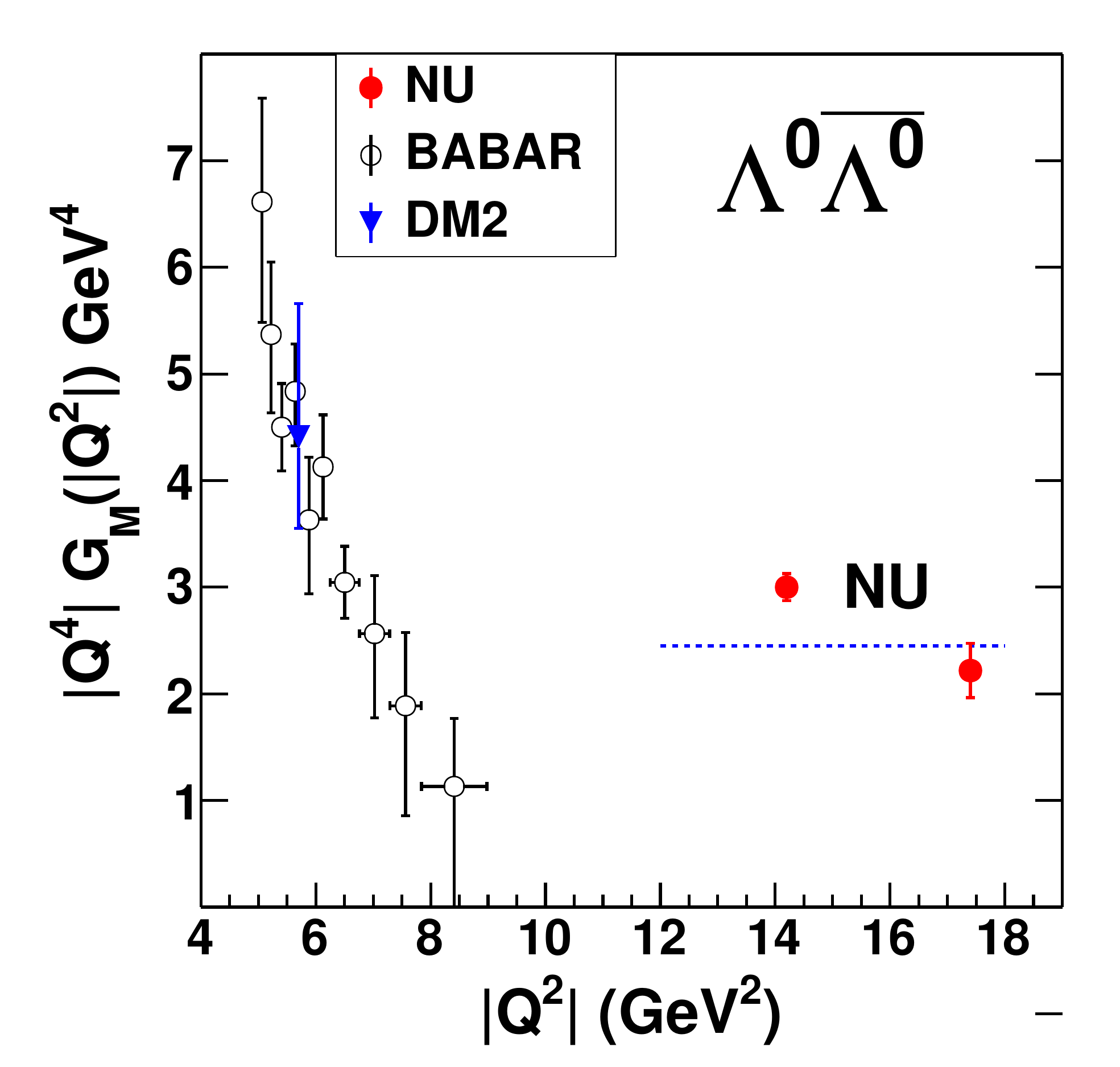}
\includegraphics[width=1.9in]{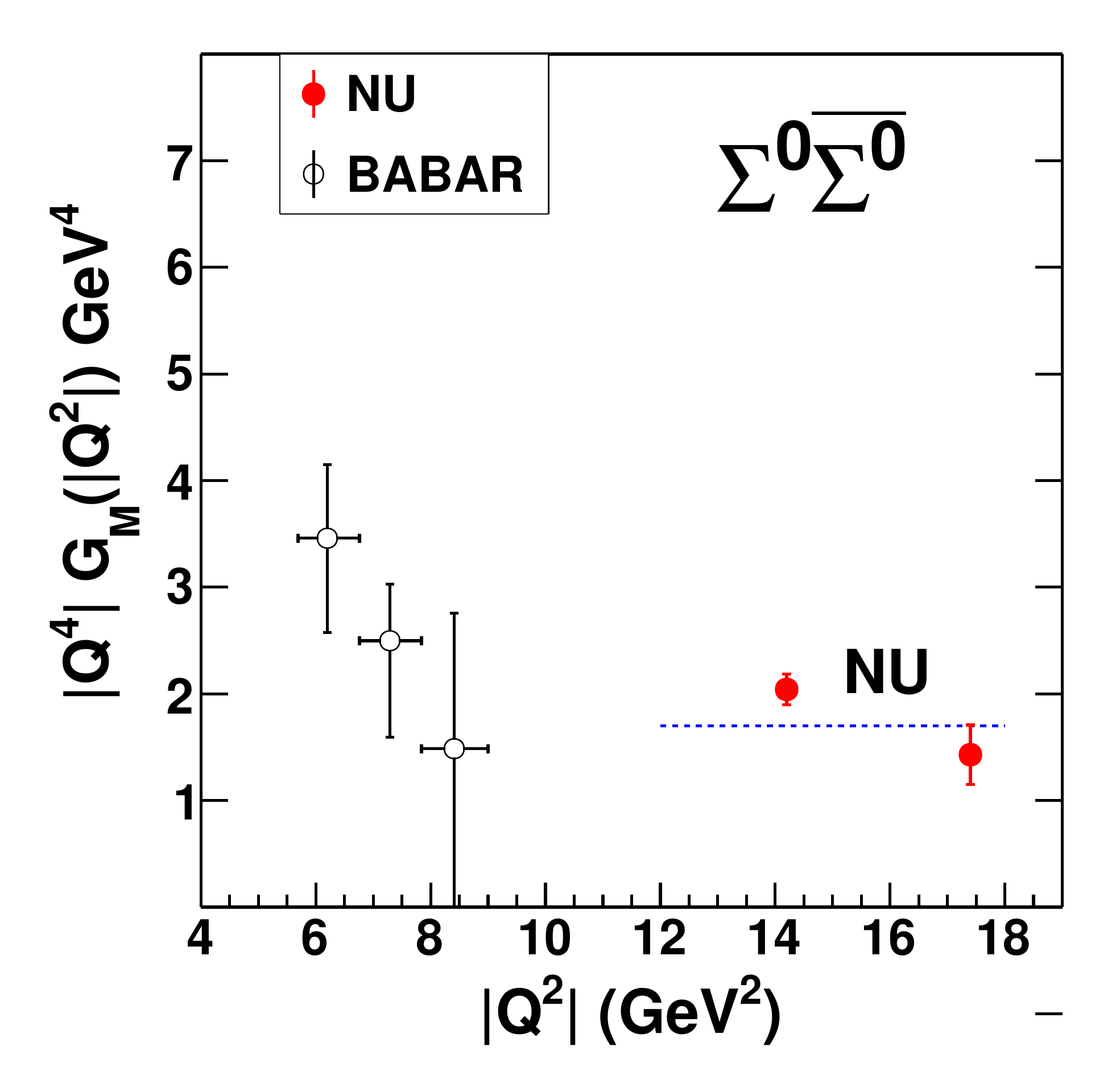}

\includegraphics[width=1.9in]{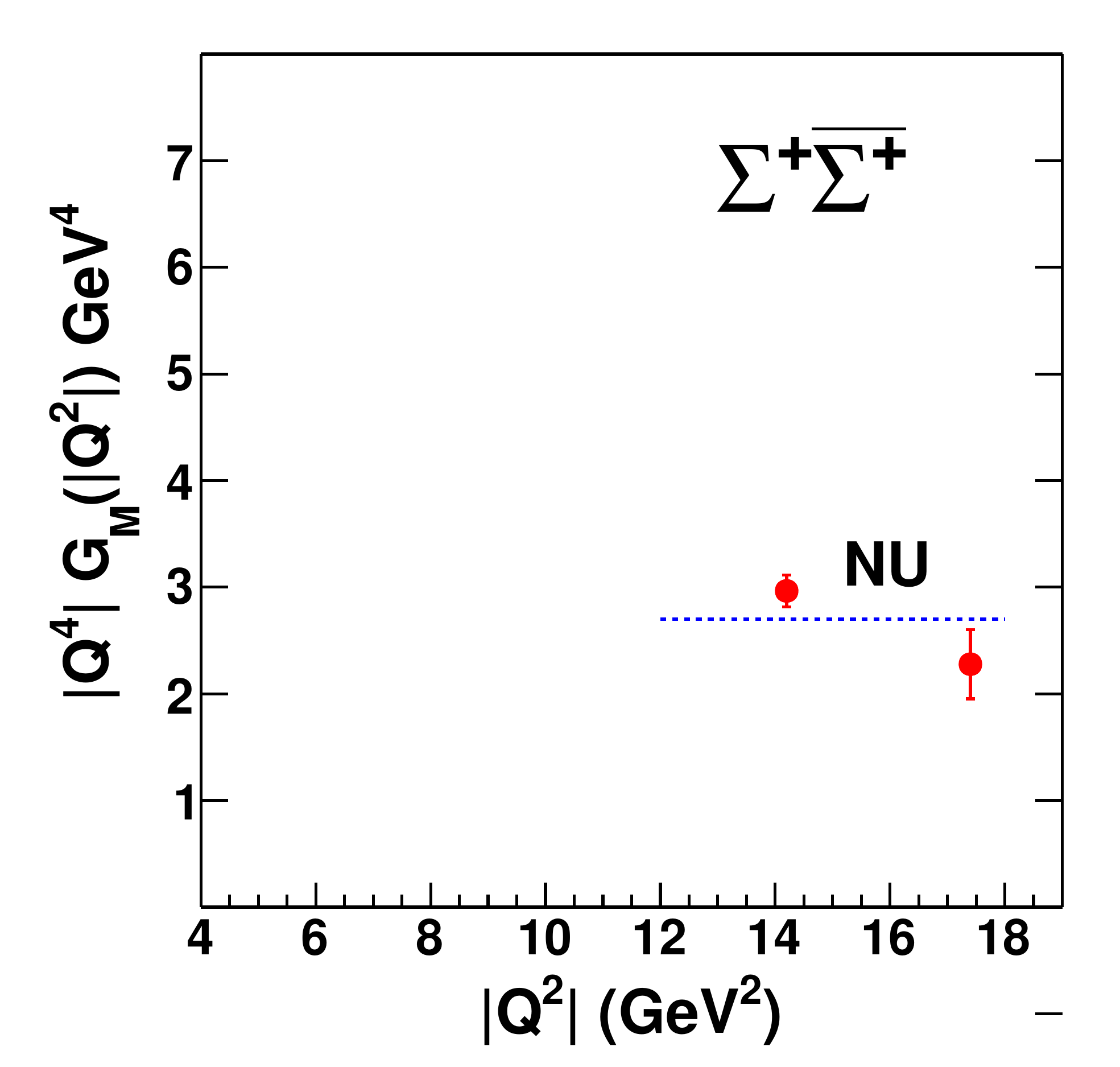}
\includegraphics[width=1.9in]{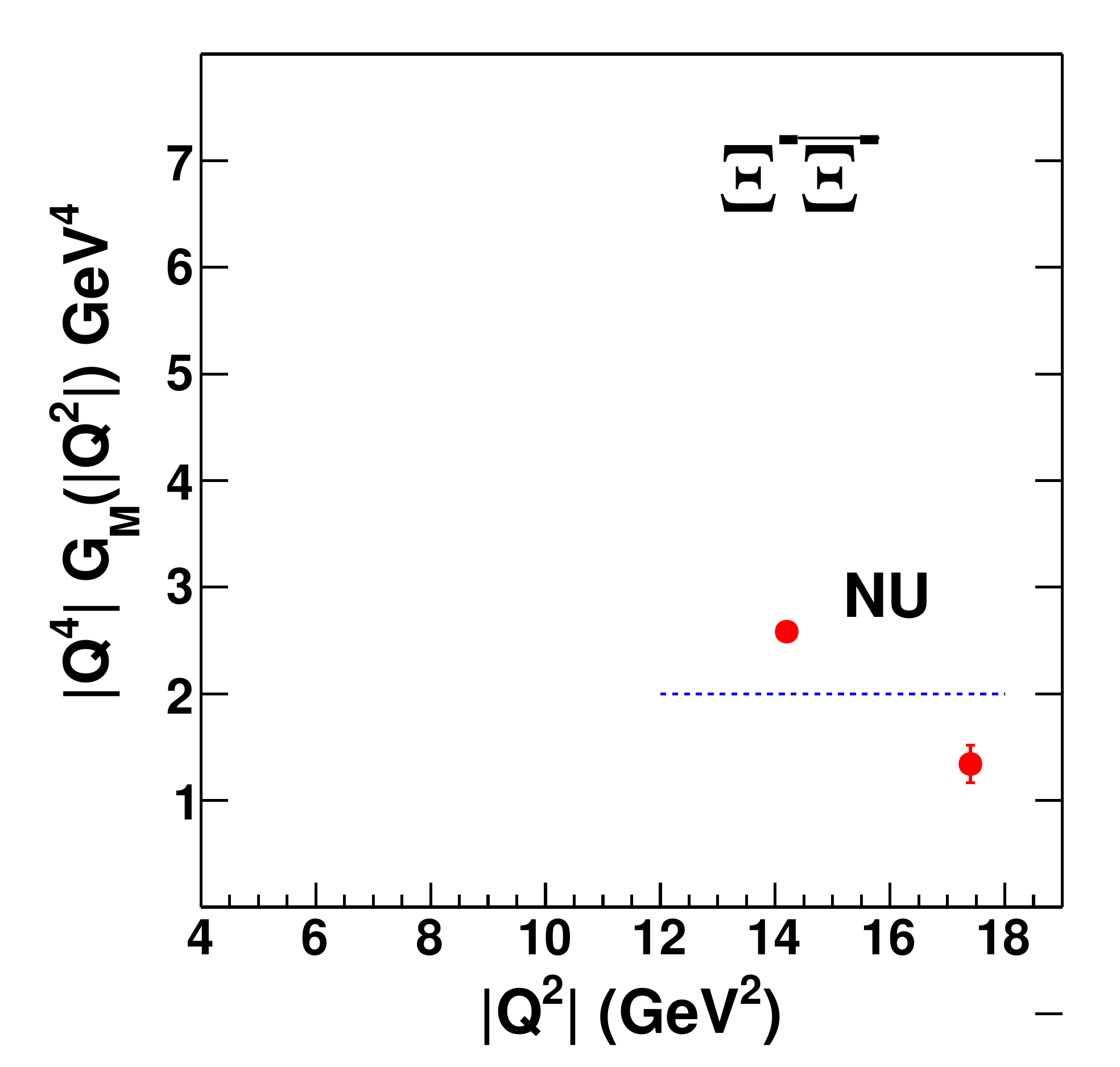}
\includegraphics[width=1.9in]{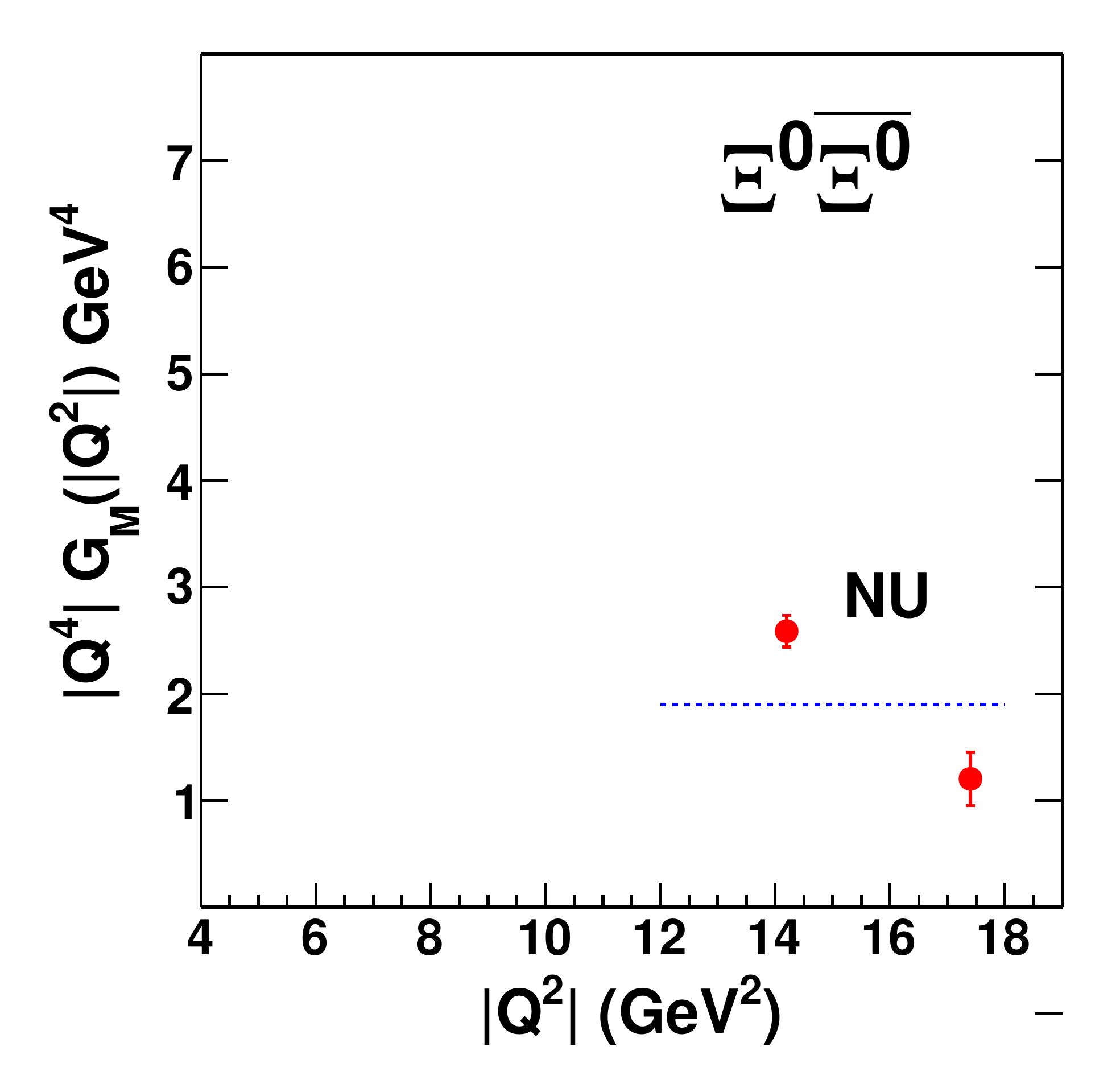}

\includegraphics[width=1.9in]{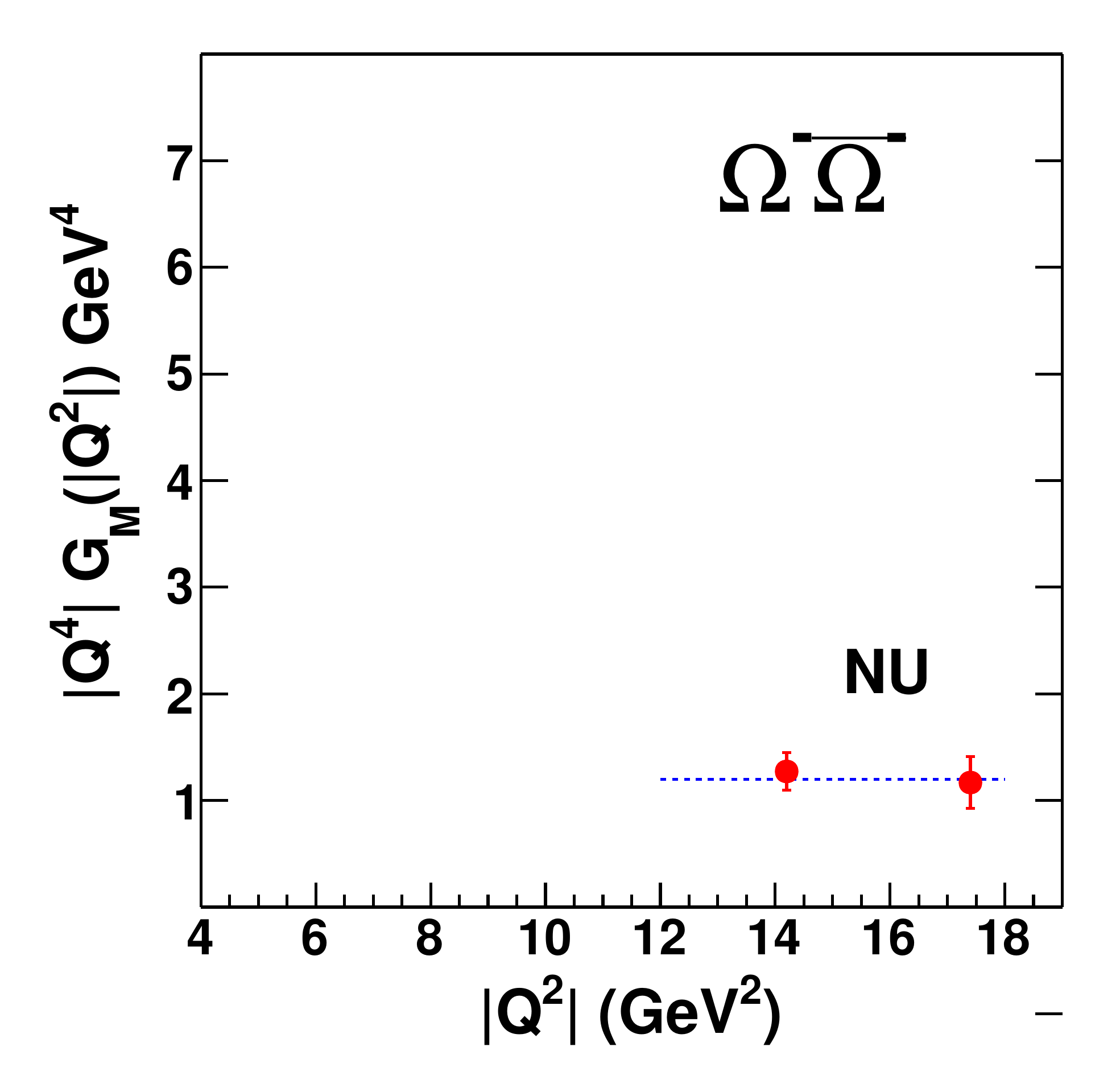}
\includegraphics[width=1.9in]{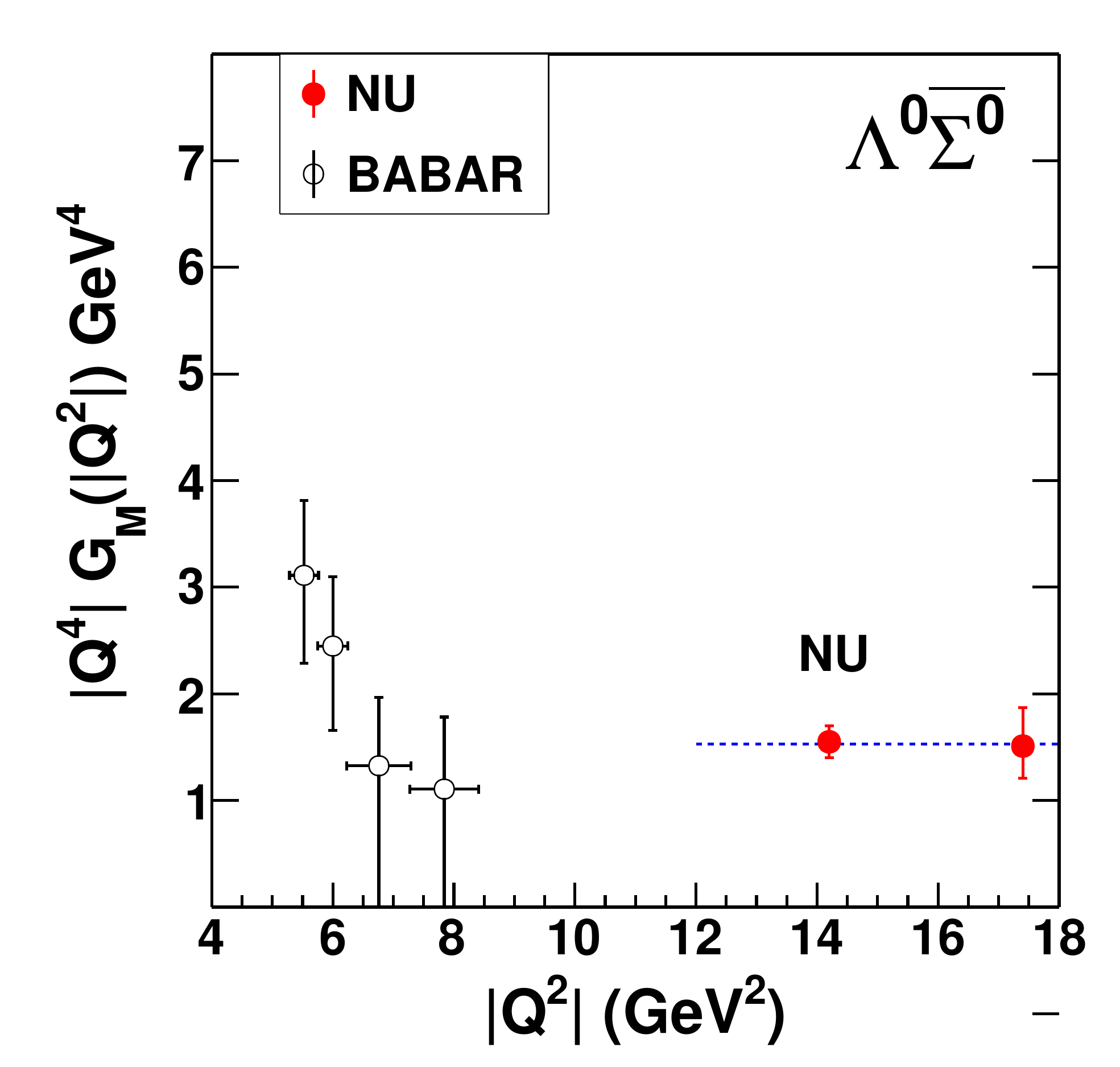}

\end{center}
\caption{Illustration of the $|Q^2|$ dependence of form factors.  Results from the present analysis are shown by the filled circles (``NU'').  Results from previous measurements by the DM2~\cite{dm2} and BaBar~\cite{babar} Collaborations are also shown with closed triangles and open circles, respectively.  The first panel shows measurements of proton timelike form factors for comparison from BaBar~\cite{babar-pp}, Fermilab E760/E835~\cite{fnal-pp}, BES~\cite{bes-pp}, and analyses of CLEO data (NU)~\cite{nu-pp}.}
\end{figure*}


\section{Summary and Discussion of Results}

We have made the world's first high precision measurements of pair production of $\Lambda^0,\Sigma^0,\Sigma^+,\Xi^0,\Xi^-$, and $\Omega^-$ hyperons at large timelike momentum transfers of $|Q^2|=13.7$, 14.2, and 17.4~GeV$^2$.
At $|Q^2|=13.7$~GeV$^2$ production is dominated by strong interaction production of the $\psi(2S)$ resonance with large cross sections. At $|Q^2|=14.2$ and 17.4~GeV$^2$ pair production is almost entirely electromagnetic, and the cross sections are smaller by orders of magnitude. No simple proportionality to the magnetic moments of the different hyperons is observed. Instead of the simple $s^5$ proportionality of the cross sections predicted by perturbative QCD, it is found that the cross sections depend on the number $n_s$ of strange quarks in the hyperons.
Quark counting rules of QCD predict a $1/s^5$ proportionality of the electromagnetic cross sections for baryons, which would lead to a constant ratio, $R = \sigma(3.77~\text{GeV})/\sigma(4.17~\text{GeV}) = 2.74$ for all hyperons. Instead, as shown in Fig.~7(c), we find that the ratio changes with the number $n_s$ of strange quarks in the hyperon, being 
$R(n_s=0,\mathrm{proton})=0.5$, $R(n_s=1,\Lambda^0,\Sigma^0,\Sigma^+)\approx 4$, and $R(n_s=2,\Xi^-,\Xi^0)\approx 10$. The spin--3/2 $\Omega^-$, and the $\Lambda^0\Sigma^0$ transition pair do not follow the trend.

The electromagnetic production data for $|Q^2|=14.2$ and 17.4~GeV$^2$ is analyzed in terms of the traditional electric and magnetic form factors, $G_E(Q^2)$ and $G_M(Q^2)$. The angular distributions of the measured cross section for $|Q^2|=14.2$~GeV$^2$ are found to be consistent with $|G_E/G_M|=0$. This rather unexpected result is at variance with BaBar's determination of $|G_E/G_M|=1$ for $\Lambda\bar{\Lambda}$ production for $|Q^2|<8$~GeV$^2$, but is in agreement with Jlab measurement of $G_E=0$ at $|Q^2|\approx 8$~GeV$^2$ for the spacelike form factor of the proton~\cite{proton_ff}.

We analyze our data for determining the timelike form factor, $G_M(Q^2)$ with the assumption $|G_E/G_M|=0$, i.e., $G_E=0$. We note however that if $|G_E/G_M|=1$ is assumed the resulting $G_M$ values would be smaller by $8-18\%$ than the values in our Tables~\ref{tbl:3770results} and \ref{tbl:4170results}, and in Figs.~8 and 9.

No pQCD or lattice-based predictions for hyperon pair production or inclusive hyperon production cross sections or timelike form factors exist.  
Two predictions based on the vector dominance (VDM) model exist. The first is the 1977 prediction of K\"orner and Kuroda~\cite{kornerkuroda} for pair production cross sections of all hyperons for $|Q^2|=$ threshold to 16 GeV$^2$. 
The other is the recent (1991) VDM calculation by Dubnickova~et~al.~\cite{dubnickova}, for the spacelike and timelike form factors from threshold to $\sqrt{s}=10$~GeV.  

No experimental data were available to  K\"orner and Kuroda in 1977 to constrain the parameters of their calculation, and their predicted cross sections at $\psi(3770)$ are found to be generally an order of magnitude smaller than our measured cross sections in Table~\ref{tbl:3770results}.  

In their VDM calculation for $\Lambda$ production Dubnickova~et~al.~\cite{dubnickova} normalize their parameters to fit the value measured by DM2 for $\Lambda$ production at $|Q^2|=5.7$~GeV$^2$.
They therefore do not designate their results for $\Lambda$ production at other energies as predictions. We note, however, that their `non-predictions' extrapolated to $|Q^2|=14.2$~GeV$^2$ giave $\sigma(\Lambda^0)=0.81$~pb, and $G_E=G_M=1.28\times10^{-2}$ in agreement with our measurements in Table~\ref{tbl:3770results}.

Our most important finding concerns the significant difference we find in the electromagnetic production cross section of $\Lambda^0$ and $\Sigma^0$ which have the same uds quark content but different isospins, and its explanation in terms of diquark correlations.

\subsection{$\bm{\Lambda^0,}$ $\bm{\Sigma^0}$ Production and Diquark Correlations}

The importance of certain configurations of flavor, spin, and isospin of two quarks in the structure of hadrons has been recognized for a long time (for a review see Anselmino \textit{et al.}~\cite{anselmino}.)
One of the best examples of the role of diquarks was provided by the Fermilab measurement of the timelike form factor of proton. It was found to be twice as large as its spacelike form factor, and it was successfully explained by considering a diquark-quark structure for the proton~\cite{kroll}.

Recently, Wilczek, Jaffe and colleagues~\cite{jaffe,wilczek,selem} have emphasized the importance of the flavor, spin, and isospin antisymmetric state of two quarks in the structure of $\Lambda^0$ and $\Sigma^0$ hyperons. Wilezek calls the spin scalar (isospin 0) diquark in $\Lambda^0$ the `good' diquark, and the spin vector (isopsin 1) diquark in $\Sigma^0$ the `bad' diquark. 
One consequence of this is that in production experiments, one expects that ``the good diquark would be significantly more likely to be produced than the bad diquark'', and that ``this would reflect itself in a large $\Lambda/\Sigma$ ratio''~\cite{wilczek}.  
Wilczek cites the LEP~\cite{pdg} observation of the relative multiplicities in the decay of $\Lambda^0$ and $\Sigma^0$, ($\sigma(\Lambda^0)/\sigma(\Sigma^0)=3.5\pm1.7$) as an important confirmation of the prediction. The decay of $\Lambda^0$ and $\Sigma^0$ in our measurements provides independent confirmation of this prediction. As listed in Tables~\ref{tbl:3770results}, and \ref{tbl:4170results}, we obtain $\sigma(\Lambda^0)/\sigma(\Sigma^0)=2.46\pm0.46$ at $\psi(3770)$, $|Q^2|=14.2$~GeV$^2$, and $\sigma(\Lambda^0)/\sigma(\Sigma^0)=2.56\pm1.40$ at $\psi(4170)$, $|Q^2|=17.4$~GeV$^2$. We consider these measurements as strong independent confirmation of the importance of diquark correlations in the structure of $\Lambda^0$ and $\Sigma^0$. Our data for the $\Xi^0$ and $\Xi^-$ containing two strange quarks should provide additional opportunity to examine other features of diquark correlations.

\begin{acknowledgments}
This investigation was done using CLEO data, and as members of the former CLEO Collaboration we thank it for this privilege.  This research was supported by the U.S. Department of Energy. 
\end{acknowledgments}

\end{document}